\documentclass[10pt, aps, prd,
reprint, 
notitlepage, nofootinbib, longbibliography, showpacs, letter, eqsecnum,
superscriptaddress,
]{revtex4-2}

\usepackage{pstricks}
\usepackage{amsmath,amssymb,bm}
\usepackage{pgfplots}
\usepackage{graphicx}
\usepackage{verbatim}
\usepackage[breaklinks=true,colorlinks=true,urlcolor=blue,linkcolor=red,citecolor=red]{hyperref}

\usepackage{bigints}
\usepackage{rotating}
\numberwithin{equation}{section}

\begin{document}

\title{Casimir-Polder energy landscape: Unipolarizable atom and ring}

\author{Niranjan Warnakulasooriya}
\email{niranjanwar@gmail.com} 
\affiliation{School of Physics and Applied Physics,
Southern Illinois University--Carbondale, Carbondale, Illinois 62901, USA}

\author{John Joseph Marchetta}
\email{jjmarchetta@gmail.com} 
\affiliation{School of Physics and Applied Physics,
Southern Illinois University--Carbondale, Carbondale, Illinois 62901, USA}

\author{Prachi Parashar}
\email{Prachi.Parashar@jalc.edu} 
\affiliation{Life and Physical Science Department,
John A. Logan College, Carterville, Illinois 62918, USA}

\author{K. V. Shajesh}
\email{kvshajesh@gmail.com}
\affiliation{School of Physics and Applied Physics,
Southern Illinois University--Carbondale, Carbondale, Illinois 62901, USA}

\date{\today}

\begin{abstract}
The Casimir-Polder interaction energy between a unipolarizable point atom
and a unipolarizable dielectric ring has been limited, until now, to
the case when the atom is confined on the axis of symmetry of the ring.
We find the generalized analytical expression for any position of the
atom relative to the ring in terms of complete elliptic integrals.
This is aided by the construction of a class of integrals of a Jacobian
elliptic function as a linear combination of complete elliptic integrals.
Our expression for the interaction energy allows us to investigate the
instability of the atom even for the equilibrium points that exist off
the axis of symmetry.
\end{abstract}

\maketitle

\section{Introduction}

The Casimir-Polder interaction is the long-range retarded effect associated with the 
Casimir effect~\cite{Casimir:1948pc},
where it is typically dominated by the zero-frequency (static) mode. The retardation leads to
the inverse seventh power law behavior, in contrast to the inverse sixth
power law decay of the nonretarded London and van der Waals interaction 
energy~\cite{Waals:1873sl,London:1930a,London:1930b,Hettema:2001cq},
becoming one of the earliest demonstrations of an interplay between quantum
and relativistic effects in a system. It was also realized early on that
these interactions could exhibit repulsive forces between polarizable 
atoms~\cite{Axilrod:1943at,Muto:1943fc,Craig:1969pa,Craig1969ma,Babb2005ac}.
More recently, in Ref.~\cite{Levin:2010vo},
it was shown that the interaction energy between an
elongated needle-shaped conductor and a perfectly conducting metal sheet with
a circular aperture could also exhibit repulsion. This led to investigations of
similar configurations in the non-retarded van der Waals regime in
Refs.~\cite{Eberlein2011hwp} and
\cite{Abrantes2018tcp}
and an analytic expression for the retarded Casimir-Polder interaction energy
between a unipolarizable atom and a dielectric plate
with a circular aperture
was derived in Refs.~\cite{Milton2012esc} and \cite{Shajesh:2011daa}. Later,
in Refs.~\cite{Marchetta:2020rps,Marchetta:2020dap}, a closed-form expression
for the retarded energy for the case of circular dielectric ring and an atom was derived.
A limitation of all of the above investigations is that the position of the atom
is restricted to the axis of symmetry of the configuration.
Even though this restriction simplifies the problem sufficiently
for the interaction energy to have a simple form in terms of rational functions,
it does not allow for any stability analysis.

Our results generalize the interaction energy
when the atom is off the axis of symmetry of the ring.
We show that the Casimir-Polder interaction energies between
a unipolarizable atom and a unipolarizable dielectric ring
admits exact solutions in terms of elliptic integrals.
In particular, we accomplish this for one of the many configurations
considered in Refs.~\cite{Marchetta:2020dap,Marchetta:2020rps}
when the atom is no more restricted to the axis of symmetry.
Even though it was realized early on that in this configuration the
unipolarizable atom has radial instability everywhere on the symmetry
axis, the analytical expression for the interaction energy we present
here allows us to explore the nature of instability concretely
in the vicinity of equilibrium points even when these points
are off the axis of symmetry.
This article is part of a broader program 
\cite{Shajesh:2011daa, Marchetta:2020rps, Marchetta:2020dap, Warnakulasooriya2023mrm}
that has been developing techniques to investigate whether a spinning
polarizable particle can be stabilized above a polarizable ring
using interactions manifested by quantum vacuum fluctuations,
an analog of a spinning magnet levitating above a magnetic ring using
classical electromagnetic interactions.

In the next section (Section~\ref{sec-CP-Energy}), we describe our system 
constituting of a unipolarizable atom and a unipolarizable dielectric
ring. We introduce the expression for
the Casimir-Polder interaction energy for the configuration
as an integral over an azimuth angle.
In Section~\ref{sec-atom-on-axis}, we review and collect the relevant results in
Ref.~\cite{Marchetta:2020rps} when the unipolarizable atom is
confined on the axis of symmetry.
In Section~\ref{sec-com-Ell-Int}, we construct a class of integrals of a Jacobian
elliptic function $\pi_n(k)$ and express them as a linear combination
of complete elliptic integrals. A recurrence relation for $\pi_n(k)$
is presented~\cite{Warnakulasooriya2023mrm}.
In terms of these elliptic integrals, in Section~\ref{sec-atom-off-axis},
we derive a generalized expression for the Casimir-Polder interaction energy
between a unipolarizable atom and a unipolarizable dielectric ring
when the atom is off the axis of the symmetry of the configuration.
We reproduce the interaction energy in Ref.~\cite{Marchetta:2020rps} 
when the atom is confined to the axis of symmetry.
In Section~\ref{sec-stab-ana}, we investigate the nature of stability of the atom
when it is placed on the equilibrium points for a couple of
orientations of the polarizability of atom.
We present an outlook and conclusion in Section~\ref{sec-conclu}.

\section{Casimir-Polder energy}
\label{sec-CP-Energy}

The Casimir-Polder interaction energy between a unipolarizable
point atom described by
its atomic polarizability ${\bm\alpha}$ and a unipolarizable
dielectric of susceptibility tensor ${\bm\chi}$ is given by
\cite{Marchetta:2020dap}
\begin{eqnarray}
&& E({\bm\alpha}, {\bm\chi}; {\bf r})
=-\frac{\hbar c}{32\pi^2} \int \frac{d^3 r^\prime}{|{\bf r}^\prime -{\bf r}|^7}
\Big[ 13\,\text{tr} ({\bm\alpha} \cdot {\bm\chi})
\hspace{15mm} \nonumber \\ && \hspace{10mm}
-56 (\hat{\bf s} \cdot {\bm\alpha} \cdot {\bm\chi} \cdot \hat{\bf s})
+63 (\hat{\bf s} \cdot {\bm\alpha} \cdot \hat{\bf s})
 (\hat{\bf s} \cdot {\bm\chi} \cdot \hat{\bf s}) \Big],
\label{f49en}
\end{eqnarray}
where the unit vector
\begin{equation}
\hat{\bf s} = \frac{{\bf r}^\prime -{\bf r}}{|{\bf r}^\prime -{\bf r}|}
\end{equation}
is constructed out of vector
\begin{equation}
{\bf s} = {\bf r}^\prime -{\bf r},
\label{rel-pos-def}
\end{equation}
which describes the relative position ${\bf r}$ of the atom
with respect to a point ${\bf r}^\prime$ on the dielectric.
Here, $\hbar$ is the rationalized Planck constant
and $c$ is speed of light in vacuum.
For the case of dielectrics with uniform susceptibility, the Casimir-Polder
interaction energy,
\begin{eqnarray}
E({\bm\alpha}, {\bm\chi}; {\bf r})
&=&-\frac{\hbar c}{32\pi^2} \,\text{tr} \Big[ 13\, ({\bm\alpha} \cdot {\bm\chi}) M_0
-56 ({\bm\alpha} \cdot {\bf M}_2 \cdot {\bm\chi})
\nonumber \\ && \hspace{16mm}
+63 ({\bm\alpha} \cdot {\bf M}_4 \cdot {\bm\chi}) \Big],
\end{eqnarray}
can be expressed in terms of a zero-rank tensor, a scalar,
\begin{equation}
M_0 = \int d^3 r^\prime \, \frac{1}{s^7},
\end{equation}
a second-rank tensor 
\begin{equation}
{\bf M}_2 = \int d^3 r^\prime \,
\frac{\hat{\bf s} \, \hat{\bf s}}{s^7},
\end{equation}
and a fourth-rank tensor
\begin{equation}
{\bf M}_4 = \int d^3 r^\prime \,
\frac{\hat{\bf s} \, \hat{\bf s} \, \hat{\bf s} \, \hat{\bf s}}{s^7}.
\end{equation}
The `dot' operations in $({\bm\alpha} \cdot {\bf M}_4 \cdot {\bm\chi})$
comprise of trace operations between second-rank tensors (that are
suitably represented as dyadics) and a fourth-rank tensor,
\begin{eqnarray}
(\hat{\bf s} \cdot {\bm\alpha} \cdot \hat{\bf s})
 (\hat{\bf s} \cdot {\bm\chi} \cdot \hat{\bf s})
&=& \text{tr}\, \big[({\bm\alpha} \cdot \hat{\bf s}) \, \hat{\bf s} \big]
\; \text{tr}\, \big[\hat{\bf s} \, (\hat{\bf s} \cdot {\bm\chi}) \big] \nonumber \\
&=& \text{tr}\,({\bm\alpha} \cdot {\bf M}_4 \cdot {\bm\chi}),
\end{eqnarray}
and is merely for typographic brevity.

\begin{figure}
\includegraphics[width=5cm]{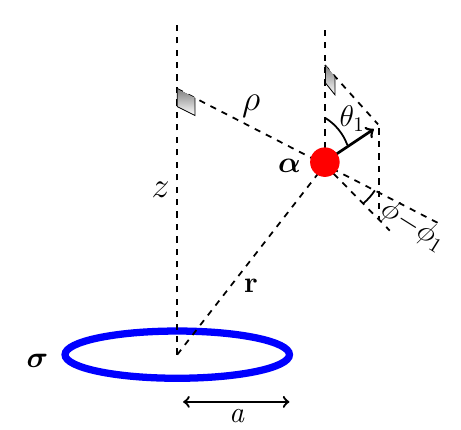}
\caption{A unipolarizable point atom with polarizability 
$\boldsymbol{\alpha}=\alpha_1\hat{\textbf{n}}\hat{\textbf{n}}$
at position ${\bf r}$ above a dielectric ring of radius $a$ with
uniform polarizability
$\boldsymbol{\sigma}=\sigma_2\hat{\textbf{z}}\hat{\textbf{z}}$.
The point atom is a distance $\rho$ away from the axis of
the ring. The principal axis $\hat{\bf n}$ of
the atomic polarizability $\boldsymbol{\alpha}$
subtends an angle $\theta_1$ with respect to polarizability of
the ring and the axis of the symmetry,
that is, $\hat{\textbf{n}} \cdot \hat{\textbf{z}}=\cos\theta_1$}.
\label{fig-aonr205}
\end{figure}

\subsection{Atom and ring}

In this article, for simplicity, we shall confine our attention to
the case of a polarizable atom that is polarizable only in a particular direction,
say $\hat{\bf n}$, such that, it's atomic polarizability can be expressed in the form
\begin{equation}
{\bm\alpha} = \alpha_1 \hat{\bf n} \hat{\bf n},
\end{equation}
where $\alpha_1$ describes the magnitude of the atomic polarizability
of the atom and $\hat{\bf n}$ is the direction of polarizability.
We will call such an atom a completely anisotropically polarizable
atom, or simply a unipolarizable atom.
Similarly, for simplicity, we shall choose the direction of polarization
of the dielectric ring to be the $\hat{\bf z}$ axis and the magnitude of
polarization to be $\sigma_2$. Thus, the ring is described by the susceptibility,
\begin{equation}
{\bm\chi} = \sigma_2 \,\hat{\bf z} \hat{\bf z} \,
\delta(z^\prime-0) \delta(\rho^\prime -a),
\end{equation}
where $a$ is the radius of the ring and coordinates $z^\prime$,
$\rho^\prime$, and $\phi^\prime$, represent the vector ${\bf r}^\prime$
that is integrated over in the expression for energy in Eq.\,(\ref{f49en}).
Then, the volume element $d^3r^\prime$ in the expression for energy is
$d^3r^\prime =ad\rho^\prime d\phi^\prime dz^\prime$
and after completing the integrations in $z^\prime$ and $\rho^\prime$,
we can express the interaction energy between the atom and ring in the form
\begin{eqnarray}
E({\bm\alpha}, {\bm\chi}; {\bf r})
=-\frac{\hbar c \alpha_1 \sigma_2}{32\pi^2} 
\int_0^{2\pi} \frac{ad\phi^\prime}{s^7}
\Big[ 13\, (\hat{\bf n} \cdot \hat{\bf z})^2
\hspace{15mm} \nonumber \\
-56 (\hat{\bf s} \cdot \hat{\bf n})
(\hat{\bf n} \cdot \hat{\bf z})
(\hat{\bf z} \cdot \hat{\bf s})
+63 (\hat{\bf n} \cdot \hat{\bf s})^2
(\hat{\bf z} \cdot \hat{\bf s})^2
\Big], \hspace{8mm}
\label{en-ex-s7205}
\end{eqnarray}
where ${\bf s}$ defined in Eq.\,(\ref{rel-pos-def}),
in the configuration under consideration
in FIG.\,\ref{fig-aonr205} is given in terms of
\begin{subequations}
\begin{eqnarray}
{\bf r} &=& \rho\,\hat{\bm\rho} +z \hat{\bf z}, \label{pacc5} \\
{\bf r}^\prime &=& a\,\hat{\bm\rho}^\prime,
\end{eqnarray}
\end{subequations}
with 
\begin{subequations}
\begin{eqnarray}
\hat{\bm\rho} &=& \cos\phi \,\hat{\bf x} + \sin\phi \,\hat{\bf y}, \\
\hat{\bm\rho}^\prime
&=& \cos\phi^\prime \hat{\bf x} + \sin\phi^\prime \hat{\bf y}.
\end{eqnarray}
\end{subequations}
Using the completeness relation,
\begin{equation}
{\bf 1} = \hat{\bm\rho} \hat{\bm\rho}
+\hat{\bm\phi} \hat{\bm\phi} +\hat{\bf z} \hat{\bf z},
\end{equation}
we can write
\begin{subequations}
\begin{eqnarray}
\hat{\bm\rho}^\prime &=&
\hat{\bm\rho}^\prime \cdot (\hat{\bm\rho} \hat{\bm\rho}
+\hat{\bm\phi} \hat{\bm\phi} +\hat{\bf z} \hat{\bf z}) \\ 
&=& (\hat{\bm\rho}^\prime \cdot \hat{\bm\rho}) \hat{\bm\rho}
 + (\hat{\bm\rho}^\prime \cdot \hat{\bm\phi}) \hat{\bm\phi}
 + (\hat{\bm\rho}^\prime \cdot \hat{\bf z}) \hat{\bf z} \\
&=& \cos(\phi^\prime-\phi) \hat{\bm\rho}
+ \sin(\phi^\prime-\phi) \hat{\bm\phi}.
\end{eqnarray}
\end{subequations}
Thus, we can show that the relative position vector ${\bf s}$
in Eq.\,(\ref{rel-pos-def}) can be written in the form
\begin{equation}
{\bf s} = [a \cos(\phi^\prime-\phi) -\rho] \,\hat{\bm\rho}
+a \sin(\phi^\prime-\phi) \,\hat{\bm\phi} -z \,\hat{\bf z},
\label{relpos-vec36}
\end{equation}
with magnitude
\begin{equation}
s = \sqrt{a^2+z^2+\rho^2-2a\rho \cos(\phi^\prime-\phi)}.
\label{relpos-mag36}
\end{equation}
Substituting $\phi^\prime-\phi \to \phi^\prime$, and using the
fact that the integral in Eq.\,(\ref{en-ex-s7205})
is fundamentally a sum that is independent of the
order of addition, we have
\begin{eqnarray}
E({\bm\alpha}, {\bm\chi}; {\bf r})
=-\frac{\hbar c \alpha_1 \sigma_2}{32\pi^2}
\int_{-\pi}^\pi \frac{ad\phi^\prime}{s^7}
\Big[ 13\, (\hat{\bf n} \cdot \hat{\bf z})^2
\hspace{15mm} \nonumber \\
-56 (\hat{\bf s} \cdot \hat{\bf n})
(\hat{\bf n} \cdot \hat{\bf z})
(\hat{\bf z} \cdot \hat{\bf s})
+63 (\hat{\bf n} \cdot \hat{\bf s})^2
(\hat{\bf z} \cdot \hat{\bf s})^2
\Big], \hspace{8mm}
\end{eqnarray}
where we suppressed the dependence on $\phi^\prime$ in 
$\hat{\bf s}(\phi^\prime)$ and $s(\phi^\prime)$, that is, 
\begin{equation}
{\bf s} (\phi^\prime) = (a \cos\phi^\prime -\rho) \,\hat{\bm\rho}
+a \sin\phi^\prime \,\hat{\bm\phi} -z \,\hat{\bf z}.
\label{spp-def}
\end{equation}
We gain some additional insight for the form of the interaction 
by expressing the energy in the form
\begin{eqnarray}
E({\bm\alpha}, {\bm\chi}; {\bf r})
=-\frac{\hbar c}{32\pi^2} \frac{\alpha_1 \sigma_2}{a^6}
\Big[ 13\, (\hat{\bf z} \cdot \hat{\bf n})^2 M_0
\hspace{20mm} \nonumber \\
-56 (\hat{\bf z} \cdot \hat{\bf n})
(\hat{\bf n} \cdot {\bf M}_2 \cdot \hat{\bf z})
+63\, \hat{\bf n} \cdot ( \hat{\bf n} \cdot {\bf M}_4 \cdot
\hat{\bf z}) \cdot \hat{\bf z}
\Big], \hspace{7mm}
\label{en-um024e}
\end{eqnarray}
where the tensors, now specific to the configuration
of atom and ring in FIG.\,\ref{fig-aonr205}, are given as
\begin{equation}
M_0 = \int_{-\pi}^\pi d \phi^\prime \, \frac{a^7}{s^7},
\label{M0-exfo45}
\end{equation}
which is a scalar,
\begin{equation}
{\bf M}_2 = a^7 \int_{-\pi}^\pi d \phi^\prime \,
\frac{\hat{\bf s} \, \hat{\bf s}}{s^7},
\label{M2-exfo45}
\end{equation}
which is a second-rank tensor written as a dyadic, and
\begin{equation}
{\bf M}_4 = a^7 \int_{-\pi}^\pi d \phi^\prime \,
\frac{\hat{\bf s} \, \hat{\bf s} \, \hat{\bf s} \, \hat{\bf s}}{s^7}
\label{M4-exfo45}
\end{equation}
is a completely symmetric fourth-rank tensor.
The order of vector operations in the construction
\begin{equation}
\hat{\bf n} \cdot ( \hat{\bf n} \cdot {\bf M}_4 \cdot
\hat{\bf z}) \cdot \hat{\bf z}
\end{equation} 
is unambiguous because of the complete symmetry in the tensor ${\bf M}_4$.
That is, the dot products contract the adjacent vectors.

Significant simplification is achieved because we confined the polarization
of the dielectric ring to be in the $\hat{\bf z}$ direction. This leads to
\begin{equation}
\hat{\bf n} \cdot {\bf M}_2 \cdot \hat{\bf z}
= -za^7 \int_{-\pi}^\pi d\phi^\prime
\frac{[\hat{\bf n} \cdot {\bf s}(\phi^\prime)]}{s^9}
\end{equation}
using $\hat{\bf z} \cdot {\bf s}(\phi^\prime)=-z$ and
\begin{equation}
\hat{\bf n} \cdot (\hat{\bf n} \cdot {\bf M}_4 \cdot
\hat{\bf z}) \cdot \hat{\bf z}
= z^2a^7 \int_{-\pi}^\pi d\phi^\prime
\frac{[\hat{\bf n} \cdot {\bf s}(\phi^\prime)]^2}{s^{11}}
\end{equation}
in terms of ${\bf s}(\phi^\prime)$ that was introduced in
Eq.\,(\ref{spp-def}).
The scalar construction appearing in the above expression is
\begin{equation}
\hat{\bf n} \cdot {\bf s}(\phi^\prime)
=(a\cos\phi^\prime -\rho) (\hat{\bf n} \cdot \hat{\bm\rho})
+a\sin\phi^\prime (\hat{\bf n} \cdot \hat{\bm\phi})
-z\, (\hat{\bf n} \cdot \hat{\bf z}),
\end{equation}
where all the $\phi^\prime$ dependence has been presented explicitly.
In particular, note that the constructions
$(\hat{\bf n} \cdot \hat{\bm\rho})$,
$(\hat{\bf n} \cdot \hat{\bm\phi})$, and
$(\hat{\bf n} \cdot \hat{\bf z})$, do not have $\phi^\prime$
dependence in them.

\section{Atom on the axis of symmetry}
\label{sec-atom-on-axis}

When the polarizable atom is positioned on the axis of symmetry of the ring,
without its direction of polarizability $\hat{\bf n}$
necessarily along the axis, we have
\begin{equation}
\rho=0,
\end{equation} 
and this leads to significant simplification in the expression
for the magnitude in Eq.\,(\ref{relpos-mag36}),
\begin{equation}
s \xrightarrow{\rho\to 0} \sqrt{a^2+z^2},
\label{smag-ax64}
\end{equation}
and in the expression for the vector in Eq.\,(\ref{relpos-vec36}),
\begin{equation}
{\bf s}(\phi^\prime) \xrightarrow{\rho\to 0} a\cos\phi^\prime \hat{\bm\rho}
+ a\sin\phi^\prime \hat{\bm\phi} - z\, \hat{\bf z}.
\label{svec-ax64}
\end{equation}
As a consequence of the above simplification, we have the following
exact evaluation of $M_0$ in Eq.\,(\ref{M0-exfo45}):
\begin{align}
M_0 \xrightarrow{\rho\to 0} 
&\; \frac{a^7}{(a^2+z^2)^\frac{7}{2}} \int_{-\pi}^\pi d\phi^\prime
\nonumber \\
= &\; 2\pi \frac{a^7}{(a^2+z^2)^\frac{7}{2}},
\end{align}
and the integral in Eq.\,(\ref{M2-exfo45}) reduces to
\begin{align}
\hat{\bf n} \cdot {\bf M}_2 \cdot \hat{\bf z}
\xrightarrow{\rho\to 0}
&\; -\frac{za^7}{(a^2+z^2)^\frac{9}{2}} \int_{-\pi}^\pi d\phi^\prime
[\hat{\bf n} \cdot {\bf s}(\phi^\prime)]
\nonumber \\
=&\; 2\pi \frac{z^2a^7 (\hat{\bf n} \cdot \hat{\bf z})}{(a^2+z^2)^\frac{9}{2}},
\label{tenM2-onax}
\end{align}
and the integral in Eq.\,(\ref{M4-exfo45}) leads to
\begin{eqnarray}
\hat{\bf n} \cdot (\hat{\bf n} \cdot {\bf M}_4 \cdot
\hat{\bf z}) \cdot \hat{\bf z} \xrightarrow{\rho\to 0}
 \frac{z^2a^7}{(a^2+z^2)^\frac{11}{2}} \int_{-\pi}^\pi d\phi^\prime
[\hat{\bf n} \cdot {\bf s}(\phi^\prime)]^2
\nonumber \\
= 2\pi \frac{z^2a^7}{(a^2+z^2)^\frac{11}{2}} \frac{1}{2}
\left[ a^2 + (2z^2-a^2) (\hat{\bf n} \cdot \hat{\bf z})^2 \right].
\hspace{6mm}
\label{m4-onax}
\end{eqnarray}
Using the above, the interaction energy in Eq.\,(\ref{en-um024e})
takes the form, for ${\bf r} =z\,\hat{\bf z}$ when the atom is on the axis,
\begin{eqnarray}
E({\bm\alpha}, {\bm\chi}; {\bf r})
\xrightarrow{\rho\to 0}
-\frac{\hbar c}{32\pi} \frac{\alpha_1 \sigma_2 a}{(a^2+z^2)^\frac{11}{2}}
\Big[ 63 a^2z^2 
\hspace{15mm} \nonumber \\
+(\hat{\bf n} \cdot \hat{\bf z})^2
(26a^4-123 a^2z^2 +40z^4) \Big]. \hspace{4mm}
\label{enax-nz-e}
\end{eqnarray}
Using the realization of $\hat{\bf n}$ in terms of spherical polar
coordinates, see FIG.\,\ref{fig-aonr205},
\begin{equation}
\hat{\bf n} = \sin\theta_1\cos\phi_1\hat{\bf x}
+ \sin\theta_1\sin\phi_1\hat{\bf y} + \cos\theta_1\hat{\bf z},
\label{pasc7}
\end{equation}
such that
\begin{equation}
(\hat{\bf n} \cdot \hat{\bf z})^2 = \cos^2\theta_1
= \frac{1+\cos2\theta_1}{2},
\end{equation}
we can rewrite the interaction energy in Eq.\,(\ref{enax-nz-e})
in the form
\begin{eqnarray}
E({\bm\alpha}, {\bm\chi}; {\bf r})
\xrightarrow{\rho\to 0}
-\frac{\hbar c}{64\pi} \frac{\alpha_1 \sigma_2 a}{(a^2+z^2)^\frac{11}{2}}
\Big[ (26 a^4 +3a^2z^2 +40 z^4) 
\nonumber \\
+\cos2\theta_1 (26a^4-123 a^2z^2 +40z^4) \Big], \hspace{10mm}
\label{enax-2th-e}
\end{eqnarray}
which reproduces the result in Ref.~\cite{Marchetta:2020dap}. 

When the atom is exactly at the center of the ring, we have
\begin{equation}
E({\bm\alpha}, {\bm\chi}; {\bf r})
\xrightarrow[z\to 0]{\rho\to 0} -\frac{13\hbar c}{16\pi} 
\frac{\alpha_1 \sigma_2}{a^6} (\hat{\bf n} \cdot \hat{\bf z})^2,
\label{at-cen-48}
\end{equation}
which is 0 for $\hat{\bf n} \perp \hat{\bf z}$.
The observation that energy is 0 both when the atom is
at the center of the ring and when the atom is at an infinite
distance away on the axis of the ring
leads to the argument that the energy must have a minimum value.
This nontrivial minimum for the energy is absent for
$\hat{\bf n} \parallel \hat{\bf z}$.
For the plate, this argument led the authors in Ref.~\cite{Levin:2010vo}
to successfully find such a minimum in their configuration. 

\section{Complete elliptic integrals}
\label{sec-com-Ell-Int}

The expression for Casimir-Polder energy when the atom is
off the axis of symmetry of the ring, that will be derived in 
Section~\ref{sec-atom-off-axis},
is aided by the construction of a new set of elliptic integrals. 
These are known integrals of a Jacobian elliptic function,
see Item 315.00 in Ref.\,\cite{Byrd:1971se},
that are not widely encountered in the literature.
In this section, we define these elliptic integrals $\pi_n(k)$
and introduce the relevant relations and identities for these
integrals that are useful for our work. 

Complete elliptic integrals of the first and second kind are defined
by the integral representations \cite{NIST:DLMF,NIST:2010fm}
\begin{subequations}
\begin{eqnarray}
K(k) &=& \int_0^{\frac{\pi}{2}}
\frac{d\psi}{\sqrt{1-k^2\sin^2\psi}}, \\
E(k) &=& \int_0^{\frac{\pi}{2}} d\psi
\sqrt{1-k^2\sin^2\psi},
\end{eqnarray}%
\label{KEcomei61}%
\end{subequations}%
respectively,
which are written here in Legendre's notation \cite{Hancock:1910fe,Byrd:1971se}.
The elliptic modulus or eccentricity $k$ will be restricted to the domain
\mbox{$0\le k<1$} in this discussion.
The complete elliptic integral of the third kind is defined as
\begin{equation}
\pi(k,\alpha) = \int_0^{\frac{\pi}{2}} \frac{d\psi}
{(1-\alpha^2\sin^2\psi) \sqrt{1-k^2\sin^2\psi}}.
\end{equation}
When the upper limit in the above three integrals does not go up to $\pi/2$,
they are called the respective elliptic integrals, dropping
the word complete. In general, any integral of the form 
\begin{equation}
\int dt\, R(t,\sqrt{P(t)})
\label{EI-def}
\end{equation}
is called an elliptic integral if
$P(t)$ is a polynomial of the third or fourth degree
and 
$R$ is a rational function in $t$ and $\sqrt{P(t)}$~\cite{Byrd:1971se}.
This blanket term is based on the fact that
it is always possible to express Eq.\,(\ref{EI-def})
linearly in terms of elementary functions
and elliptic integrals of the first, second, and third kind \cite{Byrd:1971se}.
As an illustration of this statement, we show that
the integrals of a Jacobian elliptic function,
see Item 315.00 in Ref.\,\cite{Byrd:1971se},
\begin{equation}    
\pi_n(k)=\int_0^{\frac{\pi}{2}}\frac{d\phi}{(1-k^2\sin^2\phi)^{\frac{n}{2}}},
\label{nth_order_elliptic_int}
\end{equation}
[not to be confused with the complete elliptic integral of third kind $\pi(\alpha,k)$,]
for $n=-1,1,3,5,\ldots,$
can be expressed as a linear combination of the complete
elliptic integrals of the first and second kind, that is,
\begin{equation}
\pi_n(k) = r_n(k) K(k) + s_n(k) E(k),
\label{KE-lin}
\end{equation}
where the coefficients $r_n(k)$ and $s_n(k)$ are rational functions of $k$.

The generic nature of the statement contained in the sentence containing
Eq.\,(\ref{EI-def}) is the reason for the terminology of elliptic integrals
associated with any integral that can be expressed in the form of Eq.\,(\ref{EI-def}).
We will use this idea to express the Casimir-Polder energy in terms of
complete elliptic integrals.
Complete elliptic integrals of the first and second kind provide
solutions to a wide range of problems.
For a few selected examples, see 
Refs.\,\cite{Hancock:1910fe,Byrd:1971se,Warnakulasooriya2023mrm}.
The term elliptic integral originated because an
arc length of an ellipse is expressed as an elliptic integral
and the perimeter of an ellipse is given in terms of the elliptic integral
of the second kind as
\begin{equation}
4 aE(k)
\end{equation}
in terms of the eccentricity $k$ of the ellipse and semimajor
axis $a$ of the ellipse.

In terms of the complete elliptic integrals $\pi_n(k)$
in Eq.\,(\ref{nth_order_elliptic_int}), 
when $n=1$ and $n=-1$, we obtain
\begin{subequations}
\begin{eqnarray}
\pi_{-1} &=& E(k), \\
\pi_{1} &=& K(k).
\end{eqnarray}
\end{subequations}
Power series expansions
for the complete elliptic integrals of the first and second kind,
\begin{subequations}
\begin{align}
K(k)&=\frac{\pi}{2}\sum_{n=0}^\infty\Biggl[\frac{(2n)!}{2^{2n}(n!)^2}\Biggr]^2k^{2n} \\
    &=\frac{\pi}{2}\Biggl[1+\frac{1}{4}k^2+\frac{9}{64}k^4+...\Biggr],
\label{K(k)_expansion} \\
E(k)&=\frac{\pi}{2}\sum_{n=0}^\infty\Biggl[\frac{(2n)!}{2^{2n}(n!)^2}\Biggr]^2\frac{k^{2n}}{(1-2n)} \\
    &=\frac{\pi}{2}\Biggl[1-\frac{1}{4}k^2-\frac{3}{64}k^4-...\Biggr],
\label{E(k)_expansion}
\end{align}
\end{subequations}
respectively, serve as independent definitions from the
integral representation in Eqs.\,(\ref{KEcomei61}).

Taking the derivative of $\pi_n(k)$ in Eq.\,(\ref{nth_order_elliptic_int})
with respect to its argument, we obtain
\begin{equation}
\frac{d\pi_n}{dk}
=nk\int_0^{\frac{\pi}{2}}d\phi\frac{\sin^2\phi}{(1-k^2\sin^2\phi)^{\frac{n+2}{2}}}
\label{label-1}
\end{equation}
and rewrite the equation in the following form:
\begin{equation}
\frac{d\pi_n}{dk} =-\frac{n}{k}\int_0^{\frac{\pi}{2}}d\phi
\frac{(-1+1-k^2\sin^2\phi)}{(1-k^2\sin^2\phi)^{\frac{n+2}{2}}}.
\end{equation}
Integration by parts yields
\begin{align}
\frac{d\pi_n}{dk} =\frac{n}{k}\int_0^{\frac{\pi}{2}}d\phi
&\Biggl[\frac{1}{(1-k^2\sin^2\phi)^{\frac{n+2}{2}}} \nonumber \\
& -\frac{1-k^2\sin^2\phi}{(1-k^2\sin^2\phi)^{\frac{n+2}{2}}}\Biggl].
\end{align}
Using the definition of $\pi_n$ in Eq.~(\ref{nth_order_elliptic_int}),
this can be expressed in terms of $\pi_{n}$ and $\pi_{n+2}$ as
\begin{equation}
\frac{d\pi_n}{dk}=\frac{n}{k}
\big[\pi_{n+2}-\pi_n \big].
\label{dpin_dk}
\end{equation}

Returning to Eq.~(\ref{label-1}) and using
the identity $\sin^2\psi d\psi = -\sin\psi d\cos\psi$,
we can also express the derivative in the following form:
\begin{equation}
\frac{d\pi_n}{dk}=-nk\int_0^{\frac{\pi}{2}} d\phi
\frac{\sin\phi}{(1-k^2\sin^2\phi)^{\frac{n+2}{2}}}
\frac{d \cos\phi}{d\phi}.
\end{equation}
Again, integrating by parts and observing that the boundary terms vanish,
we obtain 
\begin{eqnarray}
\frac{d\pi_n}{dk} &=& nk \int_0^{\frac{\pi}{2}} d\phi
\frac{\cos^2\phi}{(1-k^2\sin^2\phi)^{\frac{n+2}{2}}} \nonumber \\
&& +n(n+2)k^3\int_0^{\frac{\pi}{2}} d\phi
\frac{\sin^2\phi\cos^2\phi}{(1-k^2\sin^2\phi)^{\frac{n+4}{2}}}.
\hspace{6mm}
\end{eqnarray}
Using $\cos^2\phi = 1-\sin^2\phi$, 
multiplying and dividing by $(1-k^2\sin^2\phi)$ in the first integral,
and simplifying to combine like terms, we have
\begin{eqnarray}
\frac{d\pi_n}{dk} &=& \int_0^{\frac{\pi}{2}}d\phi
\Biggl[\frac{nk}{(1-k^2\sin^2\phi)^{\frac{n+4}{2}}} \nonumber \\ &&
+\frac{[n(n+1)k^3-nk]\sin^2\phi}{(1-k^2\sin^2\phi)^{\frac{n+4}{2}}}
\nonumber \\ &&
-\frac{n(n+1)k^3\sin^4\phi}{(1-k^2\sin^2\phi)^{\frac{n+4}{2}}} \Biggr].
\end{eqnarray}
We rewrite this in the following form:
\begin{equation}
\frac{d\pi_n}{dk}= nk\,I_1+[n(n+1)k^3-nk]\,I_2 -n(n+1)k^3\,I_3,
\label{dpi_dk}
\end{equation}
where 
\begin{subequations}
\begin{eqnarray}
I_1 &=& \int_0^{\frac{\pi}{2}} \frac{d\phi}{(1-k^2\sin^2\phi)^{\frac{n+4}{2}}}, \\
I_2 &=& \int_0^{\frac{\pi}{2}}\frac{d\phi\,\sin^2\phi}{(1-k^2\sin^2\phi)^{\frac{n+4}{2}}}, \\
I_3 &=& \int_0^{\frac{\pi}{2}}\frac{d\phi\,\sin^4\phi}{(1-k^2\sin^2\phi)^{\frac{n+4}{2}}}.
\label{I3}
\end{eqnarray}
\end{subequations}
From the definition of $\pi_n$, $I_1$ can be immediately identified as
\begin{equation}
I_1=\pi_{n+4}.
\label{I1final}
\end{equation}
We rewrite the integral $I_2$ in the form
\begin{equation}
I_2=-\frac{1}{k^2}\int_0^{\frac{\pi}{2}} d\phi
\frac{(-1+1-k^2\sin^2\phi)}{{(1-k^2\sin^2\phi)^{\frac{n+4}{2}}}}
\end{equation}
and split it into two integrals as
\begin{align}
I_2=\frac{1}{k^2}\int_0^{\frac{\pi}{2}} d\phi
&\Biggl[\frac{1}{{(1-k^2\sin^2\phi)^{\frac{n+4}{2}}}} 
\nonumber \\ &
-\frac{1-k^2\sin^2\phi}{{(1-k^2\sin^2\phi)^{\frac{n+4}{2}}}}\Biggr].
\end{align}
By simplifying the second term of the integrand
and using the definition of $\pi_n$, $I_2$ is expressed as 
\begin{equation}
I_2=\frac{1}{k^2} \big[ \pi_{n+4}-\pi_{n+2} \big].
\label{I2final}
\end{equation}
Similarly,
the numerator of the integrand in $I_3$ is written in the following form:
\begin{equation}
\sin^4\phi=\alpha(1-k^2\sin^2\phi)^2+\beta(1-k^2\sin^2\phi)+\gamma,
\label{sin4-1}
\end{equation}
where $\alpha$, $\beta$, and $\gamma$, are arbitrary constants.
By equating coefficients on both sides of the equation, these can be evaluated to be 
\begin{equation}
\alpha=\frac{1}{k^4}, \quad \beta=-\frac{2}{k^4},
\quad \text{and} \quad \gamma=\frac{1}{k^4}.
\label{constants}
\end{equation}
Using Eqs.\,(\ref{sin4-1}) and (\ref{constants}),
the integrand in Eq.~(\ref{I3}) can be written in the form
\begin{eqnarray}
I_3 &=&\frac{1}{k^4}\int_0^{\frac{\pi}{2}} d\phi\Biggl[
\frac{(1-k^2\sin^2\phi)^2}{(1-k^2\sin^2\phi)^{\frac{n+4}{2}}}
\nonumber \\&&
-\frac{2(1-k^2\sin^2\phi)}{(1-k^2\sin^2\phi)^{\frac{n+4}{2}}}
+\frac{1}{(1-k^2\sin^2\phi)^{\frac{n+4}{2}}}\Biggr].
\hspace{8mm}
\end{eqnarray}
Then, canceling the common factors from the numerator and denominator
in the first and second terms of the integrand,
and using the definition of $\pi_n$, integral $I_3$ can be written as 
\begin{equation}
I_3=\frac{1}{k^4} \big[ \pi_n-2\pi_{n+2}+\pi_{n+4} \big].
\label{I3final}
\end{equation}
Using Eqs.\,(\ref{I1final}), (\ref{I2final}), and (\ref{I3final})
in Eq.\,(\ref{dpi_dk}), we have 
\begin{eqnarray}
\frac{d\pi_n}{dk} &=&
\pi_{n+4} \frac{(n^2+2n)(k^2-1)}{k} \nonumber \\&&
-\pi_{n+2} \frac{[n^2(k^2-2)+n(k^2-3)]}{k} \nonumber \\&&
-\pi_n \frac{n(n+1)}{k}.
\label{df_dk_final}
\end{eqnarray}

Equating right-hand sides of Eqs.\,(\ref{df_dk_final}) and (\ref{dpin_dk}),
and solving for $\pi_{n+4}$, we obtain a recurrence relation
for the integrals of a Jacobian elliptic function,
see Item 315.06 in Ref.\,\cite{Byrd:1971se},
\begin{equation}
\pi_{n+4}=\frac{(n+1)(k^2-2)\pi_{n+2}+n\pi_n}{(n+2)(k^2-1)}.
\end{equation}
We can rewrite the above recurrence relation as
\begin{equation}
\pi_n=\frac{(n-3)(2-k^2)\pi_{n-2}-(n-4)\pi_{n-4}}{(n-2)(1-k^2)}, 
\label{recurrence_relation}
\end{equation}
where $n=-1,1,3,5,...$, by replacing $n$ with $n-4$.
The recurrence relation in Eq.\,(\ref{recurrence_relation}) allows us to obtain $\pi_n$ 
in terms of $\pi_{n-2}$ and $\pi_{n-4}$.
In this article, we will need $\pi_n$ for
$n=3, 5, 7, 9,$ and $11$. Thus, from Eq.\,(\ref{recurrence_relation}), we obtain
\begin{subequations}
\begin{eqnarray}
\pi_3(k)&=&\frac{\pi_{-1}}{1-k^2},
\label{pi_3_recurrence} \\
\pi_5(k)&=&\frac{2(2-k^2)\pi_{3}-\pi_1}{3(1-k^2)},
\label{pi_5_recurrence} \\
\pi_7(k)&=&\frac{4(2-k^2)\pi_{5}-3\pi_3}{5(1-k^2)},
\label{pi_7_recurrence} \\
\pi_9(k)&=&\frac{6(2-k^2)\pi_7-5\pi_5}{7(1-k^2)},
\label{pi_9_recurrence} \\
\pi_{11}(k)&=&\frac{8(2-k^2)\pi_{9}-7\pi_7}{9(1-k^2)}.
\label{pi_11_recurrence}
\end{eqnarray}%
\label{piVEin}%
\end{subequations}%

As proposed in Eq.\,(\ref{KE-lin}), our goal in this Section
is to write each expression in Eqs.\,(\ref{piVEin}) as
a linear combination of $K(k)$ and $E(k)$.
To this end, first we replace $\pi_{-1}$ with $E(k)$
in Eq.\,(\ref{pi_3_recurrence}) to obtain 
\begin{equation}
\pi_3(k)=\frac{E(k)}{(1-k^2)}.
\label{pi_3_from_Ek}
\end{equation}
Using the resultant $\pi_3(k)$ and replacing $\pi_1(k)$ with $K(k)$,
we rewrite Eq.~(\ref{pi_5_recurrence}) as
\begin{equation}
\pi_5(k)=\frac{2(2-k^2)}{3(1-k^2)^2}E(k)-\frac{K(k)}{3(1-k^2)}.
\label{pi_5_from_Ek_Kk}
\end{equation}
This process can be continued to rewrite Eqs.~(\ref{pi_7_recurrence}),
(\ref{pi_9_recurrence}), and (\ref{pi_11_recurrence}) in terms of $K(k)$ and $E(k)$.
From these expressions, in terms of factor
\begin{equation}
q = 1-k^2,
\end{equation}
we can read out the rational functions $r_n(k)$ and $s_n(k)$,
\begin{subequations}
\begin{eqnarray}
r_3(k) &=& 0, \\
r_5(k) &=&
-\frac{1}{3}\frac{1}{q}, \\
r_7(k) &=&
-\frac{4}{15}\frac{1}{q^2}
-\frac{4}{15}\frac{1}{q}, \\
r_9(k) &=& 
-\frac{8}{35}\frac{1}{q^3}
-\frac{23}{105}\frac{1}{q^2}
-\frac{8}{35}\frac{1}{q}, \\
r_{11}(k) &=&
-\frac{64}{315}\frac{1}{q^4}
-\frac{4}{21}\frac{1}{q^3}
-\frac{4}{21}\frac{1}{q^2}
-\frac{64}{315}\frac{1}{q},
\hspace{5mm}
\end{eqnarray}
\end{subequations}
and
\begin{subequations}
\begin{eqnarray}
s_3(k) &=& \frac{1}{q}, \\
s_5(k) &=&
\frac{2}{3}\frac{1}{q^2}
+\frac{2}{3}\frac{1}{q}, \\
s_7(k) &=&
\frac{8}{15}\frac{1}{q^3}
+\frac{7}{15}\frac{1}{q^2}
+\frac{8}{15}\frac{1}{q}, \\
s_9(k) &=&  
\frac{16}{35}\frac{1}{q^4}
+\frac{8}{21}\frac{1}{q^3}
+\frac{8}{21}\frac{1}{q^2} 
+\frac{16}{35}\frac{1}{q}, \\
s_{11}(k) &=& 
\frac{128}{315}\frac{1}{q^5}
\frac{104}{315}\frac{1}{q^4}
+\frac{11}{35}\frac{1}{q^3} 
+\frac{104}{315}\frac{1}{q^2}
+\frac{128}{315}\frac{1}{q}.
\hspace{11mm}
\end{eqnarray}
\end{subequations}
Thus, we have successfully expressed $\pi_n(k)$
as a linear combination of $K(k)$ and $E(k)$
for $n=3,5,7,9,11$.

In addition, using the power series expansions of $E(k)$ and $K(k)$,
along with power series expansions of $1/(1 - k^2)^{j}$
for $j = 1,2,3,\ldots$, power series expansions
for $\pi_n(k)$ are obtained. They are,
\begin{subequations}
\begin{eqnarray}
\pi_3(k)&=&\frac{\pi}{2}\biggl[1+\frac{3}{4}k^2+\frac{45}{64}k^4+...\biggr],\\
\pi_5(k)&=&\frac{\pi}{2}\biggl[1+\frac{5}{4}k^2+\frac{105}{64}k^4+...\biggr],
\label{Eq_pi5_powerSeriesExpansion} \\
\pi_7(k)&=&\frac{\pi}{2}\biggl[1+\frac{7}{4}k^2+\frac{189}{64}k^4+...\biggr],
\label{Eq_pi7_powerSeriesExpansion} \\
\pi_9(k)&=&\frac{\pi}{2}\biggl[1+\frac{9}{4}k^2+\frac{297}{64}k^4+...\biggr],
\label{Eq_pi9_powerSeriesExpansion} \\
\pi_{11}(k)&=&\frac{\pi}{2}\biggl[1+\frac{11}{4}k^2+\frac{429}{64}k^4+...\biggr].
\label{Eq_pi11_powerSeriesExpansion}
\end{eqnarray}%
\label{piN_powerSeriesExpansions}%
\end{subequations}%
These expansions will be useful when we try to match our results
on the axis of symmetry, obtained by the approximation $k\ll 1$.

\section{Atom off the axis of symmetry}
\label{sec-atom-off-axis}

If the atom is not on the axis of symmetry,
the distance $s(\phi^\prime)$ in Eq.\,(\ref{relpos-mag36})
is dependent on the integration variable $\phi^\prime$,
unlike the simplification in Eq.\,(\ref{smag-ax64}),
and we have to work with the full
$\phi^\prime$ dependence in the distance. 
Nevertheless, we shall show that the interaction energy
can be expressed using elliptic integrals.

The scalar $M_0$ in Eq.\,(\ref{M0-exfo45})
involves only the distance $s(\phi^\prime)$.
We go through a sequence of straightforward substitutions
that allows us to recognize the integral as a complete elliptic 
integral. 
We begin by exploiting the symmetry of the integrand about 
$\phi^\prime=0$ to write
\begin{equation}
M_0= 2\int_0^\pi d\phi^\prime
\frac{a^7}{(a^2+\rho^2+z^2-2a\rho\cos\phi^\prime)^\frac{7}{2}}.
\end{equation}
Then, substituting $\phi^\prime\to\pi-\phi^\prime$, we have
\begin{equation}
M_0= 2\int_0^\pi d\phi^\prime
\frac{a^7}{(a^2+\rho^2+z^2+2a\rho\cos\phi^\prime)^\frac{7}{2}}.
\end{equation}
Finally, substituting $\phi^\prime=2\psi$, and using
$\cos2\psi=1-2\sin^2\psi$, we have
\begin{equation}
M_0= 4\int_0^\frac{\pi}{2} d\psi
\frac{a^7}{((a+\rho)^2+z^2-4a\rho\sin^2\psi)^\frac{7}{2}}.
\end{equation}
In terms of
\begin{equation}
k^2 = \frac{4a\rho}{(a+\rho)^2 +z^2},
\label{k2def-48}
\end{equation}
we can write
\begin{equation}
M_0 = 4 \left( \frac{a^2k^2}{4a\rho} \right)^\frac{7}{2}
 \int_0^\frac{\pi}{2} d\psi \frac{1}{(1-k^2\sin^2\psi)^\frac{7}{2}}.
\end{equation}
Thus, we have
\begin{eqnarray}
M_0 &=& 4 \left( \frac{a^2k^2}{4a\rho} \right)^\frac{7}{2} \pi_7(k).
\label{m0co-840}
\end{eqnarray}

Repeating the same sequence of substitutions for the dyadic ${\bf M}_2$
in Eq.\,(\ref{M2-exfo45}), we can show that
\begin{eqnarray}
{\bf M}_2= 2 \left( \frac{a^2k^2}{4a\rho} \right)^\frac{9}{2}
\frac{1}{a^2} \int_0^\frac{\pi}{2}
\frac{d\psi}{(1-k^2\sin^2\psi)^\frac{9}{2}}
\hspace{15mm} \nonumber \\ \times \Big[
{\bf s}(\pi-2\psi) {\bf s}(\pi-2\psi) {\bf s}(-\pi+2\psi) {\bf s}(-\pi+2\psi)
\Big]. \hspace{5mm}
\end{eqnarray}
Interaction energy involves
\begin{eqnarray}
\hat{\bf n} \cdot {\bf M}_2 \cdot \hat{\bf z}
&=& -2 \left( \frac{a^2k^2}{4a\rho} \right)^\frac{9}{2} \frac{z}{a}
\frac{1}{a} \int_0^\frac{\pi}{2}
\frac{d\psi}{(1-k^2\sin^2\psi)^\frac{9}{2}}
\nonumber \\ && \times \Big[
\hat{\bf n} \cdot {\bf s}(\pi-2\psi) +\hat{\bf n} \cdot {\bf s}(-\pi+2\psi)
\Big], \hspace{6mm}
\end{eqnarray}
where we used
\begin{equation}
\hat{\bf z} \cdot {\bf s}(\pi-2\psi)
=\hat{\bf z} \cdot {\bf s}(-\pi+2\psi) = -z.
\end{equation}
We evaluate the numerator and suitably express it in the form
\begin{eqnarray}
\frac{\hat{\bf n} \cdot {\bf s}(\pi-2\psi) 
+\hat{\bf n} \cdot {\bf s}(-\pi+2\psi)}{a}
\hspace{25mm} \nonumber \\
= -2 (\hat{\bf n} \cdot \hat{\bm\rho})
\left( 1 +\frac{\rho}{a} -\frac{2}{k^2} \right)
-2 \frac{z}{a} (\hat{\bf n} \cdot \hat{\bf z})
\nonumber \\
-\frac{4}{k^2} (\hat{\bf n} \cdot \hat{\bm\rho}) (1-k^2\sin^2\psi),
\hspace{16mm}
\end{eqnarray}
which allows us to write
\begin{eqnarray}
\hat{\bf n} \cdot {\bf M}_2 \cdot \hat{\bf z}
= 4 \left( \frac{a^2k^2}{4a\rho} \right)^\frac{9}{2} \pi_9(k)
 \frac{z^2}{a^2} (\hat{\bf n} \cdot \hat{\bf z}) 
+ 4 \left( \frac{a^2k^2}{4a\rho} \right)^\frac{9}{2}
\hspace{5mm} \nonumber \\ \times 
\left( \left( 1 +\frac{\rho}{a} \right) \pi_9(k)
-\frac{2}{k^2} \left\{ \pi_9(k)-\pi_7(k) \right\} \right)
\frac{z}{a} (\hat{\bf n} \cdot \hat{\bm\rho})
\hspace{9mm}
\label{m2co-840}
\end{eqnarray}
in terms of elliptic integrals.
In the limit $\rho\to 0$, the term with $(\hat{\bf n} \cdot \hat{\bm\rho})$
goes to zero using
\begin{equation}
\frac{a^2k^2}{4a\rho} =\frac{a^2}{a^2+z^2} 
+{\cal O}\left( \frac{\rho}{a} \right)
\end{equation}
and
\begin{eqnarray}
\left( \pi_9(k) -\frac{2}{k^2} \Big\{ \pi_9(k)-\pi_7(k) \Big\} \right)
\hspace{20mm} \nonumber \\
= \frac{\pi}{2} \left[ 0 -\frac{9}{8} k^2 
-\frac{99}{32} k^4 +\ldots \right]. \hspace{5mm}
\end{eqnarray}
Using these limits in Eq.\,(\ref{m2co-840}), we obtain
\begin{equation}
\hat{\bf n} \cdot {\bf M}_2 \cdot \hat{\bf z}
\xrightarrow{\rho\to 0}
2\pi \frac{z^2a^7 (\hat{\bf n} \cdot \hat{\bf z})}{(a^2+z^2)^\frac{9}{2}},
\end{equation}
which reproduces the evaluation in Eq.\,(\ref{tenM2-onax}).

For the term involving ${\bf M}_4$, we can show that
\begin{eqnarray}
\hat{\bf n} \cdot (\hat{\bf n} \cdot {\bf M}_4 \cdot \hat{\bf z}) \cdot \hat{\bf z}
= 2 \left( \frac{a^2k^2}{4a\rho} \right)^\frac{11}{2} \frac{z^2}{a^2}
\frac{1}{a^2} \int_0^\frac{\pi}{2}
\frac{d\psi}{(1-k^2\sin^2\psi)^\frac{9}{2}}
\nonumber \\ \times
\Big[
[\hat{\bf n} \cdot {\bf s}(\pi-2\psi)]^2 +[\hat{\bf n} \cdot {\bf s}(-\pi+2\psi)]^2
\Big]. \hspace{13mm}
\end{eqnarray}
To recognize the associated elliptic integrals, we write
\begin{eqnarray}
&& \frac{[\hat{\bf n} \cdot {\bf s}(\pi-2\psi)]^2 
+[\hat{\bf n} \cdot {\bf s}(-\pi+2\psi)]^2}{a^2}
\nonumber \\
&=& 2 \left[ (\hat{\bf n} \cdot \hat{\bm\rho})
\left( 1 +\frac{\rho}{a} -\frac{2}{k^2} \right)
+ \frac{z}{a} (\hat{\bf n} \cdot \hat{\bf z}) \right]^2
\nonumber \\ &&
+\frac{8}{k^2} \left( 1-\frac{1}{k^2} \right) (\hat{\bf n} \cdot \hat{\bm\phi})^2
+\frac{8}{k^2} \Bigg[
(\hat{\bf n} \cdot \hat{\bm\rho})^2 
\left( 1 +\frac{\rho}{a} -\frac{2}{k^2} \right)
\nonumber \\ &&
+\frac{z}{a} (\hat{\bf n} \cdot \hat{\bm\rho}) (\hat{\bf n} \cdot \hat{\bf z})
-(\hat{\bf n} \cdot \hat{\bm\phi})^2 \left( 1-\frac{2}{k^2} \right)
\Bigg] (1-k^2\sin^2\psi)
\nonumber \\ &&
+\frac{8}{k^2} \left[ (\hat{\bf n} \cdot \hat{\bm\rho})^2 
-(\hat{\bf n} \cdot \hat{\bm\phi})^2 \right] (1-k^2\sin^2\psi)^2.
\end{eqnarray}
This leads to
\begin{eqnarray}
&&\hat{\bf n} \cdot (\hat{\bf n} \cdot {\bf M}_4 \cdot \hat{\bf z}) \cdot \hat{\bf z}
= 4 \left( \frac{a^2k^2}{4a\rho} \right)^\frac{11}{2} \frac{z^2}{a^2}
\nonumber \\ &&
\times \Bigg[ (\hat{\bf n} \cdot \hat{\bm\rho})^2 \Bigg\{
\left( 1 +\frac{\rho}{a} -\frac{2}{k^2} \right)^2 \pi_{11}(k)
\nonumber \\ && \hspace{20mm}
+\frac{4}{k^2}\left( 1 +\frac{\rho}{a} -\frac{2}{k^2} \right) \pi_9(k)
+\frac{4}{k^4} \pi_7(k) \Bigg\} 
\hspace{5mm} \nonumber \\ && \hspace{5mm}
+(\hat{\bf n} \cdot \hat{\bm\phi})^2 \Bigg\{
\frac{4}{k^2} \left( 1 -\frac{1}{k^2} \right) \pi_{11}(k)
\hspace{5mm} \nonumber \\ && \hspace{20mm}
-\frac{4}{k^2}\left( 1 -\frac{2}{k^2} \right) \pi_9(k)
-\frac{4}{k^4} \pi_7(k) \Bigg\} 
\hspace{5mm} \nonumber \\ && \hspace{5mm}
+2(\hat{\bf n} \cdot \hat{\bm\rho}) (\hat{\bf n} \cdot \hat{\bf z}) \frac{z}{a}
\left\{ \left( 1 +\frac{\rho}{a} -\frac{2}{k^2} \right) \pi_{11}(k)
+\frac{2}{k^2} \pi_9(k) \right\} 
\hspace{5mm} \nonumber \\ && \hspace{5mm}
+(\hat{\bf n} \cdot \hat{\bf z})^2 \frac{z^2}{a^2} \pi_{11}(k) \Bigg].
\label{m4co-840}
\end{eqnarray}
Using the series expansions
\begin{subequations}
\begin{eqnarray}
\left( 1 -\frac{2}{k^2} \right)^2 \pi_{11}(k)
+\frac{4}{k^2}\left( 1-\frac{2}{k^2} \right) \pi_9(k)
+\frac{4}{k^4} \pi_7(k) 
\hspace{5mm} \nonumber \\
= \frac{\pi}{2} \left[
\frac{1}{2} +\frac{11}{8} k^2 +\frac{1001}{256} k^4 +\dots \right], \hspace{10mm}
\label{nhDrh-se91}
\end{eqnarray}
\begin{eqnarray}
\frac{4}{k^2}\left( 1-\frac{1}{k^2} \right) \pi_{11}(k)
-\frac{4}{k^2}\left( 1-\frac{2}{k^2} \right) \pi_9(k)
-\frac{4}{k^4} \pi_7(k) 
\nonumber \\
= \frac{\pi}{2} \left[
\frac{1}{2} +\frac{11}{8} k^2 +\frac{715}{256} k^4 +\dots \right], \hspace{12mm}
\end{eqnarray}
and
\begin{eqnarray}
\left( \pi_{11}(k) -\frac{2}{k^2} \Big\{ \pi_{11}(k)-\pi_9(k) \Big\} \right)
\hspace{20mm} \nonumber \\
= \frac{\pi}{2} \left[ 0 -\frac{11}{8} k^2
-\frac{143}{32} k^4 +\ldots \right], \hspace{10mm}
\label{nhDrh-te912}
\end{eqnarray}
\end{subequations}
we have
\begin{eqnarray}
\hat{\bf n} \cdot (\hat{\bf n} \cdot {\bf M}_4 \cdot \hat{\bf z}) \cdot \hat{\bf z}
\xrightarrow{\rho\to 0}
4 \left( \frac{a^2k^2}{4a\rho} \right)^\frac{11}{2} \frac{z^2}{a^2}
\frac{\pi}{2} 
\hspace{15mm} \nonumber \\ \times 
\left[ (\hat{\bf n} \cdot \hat{\bf z})^2 \frac{z^2}{a^2}
+(\hat{\bf n} \cdot \hat{\bm\rho})^2 \frac{1}{2}
+(\hat{\bf n} \cdot \hat{\bm\phi})^2 \frac{1}{2} \right]. \hspace{8mm}
\end{eqnarray}
Then, using
\begin{equation}
(\hat{\bf n} \cdot \hat{\bm\rho})^2
+(\hat{\bf n} \cdot \hat{\bm\phi})^2
+(\hat{\bf n} \cdot \hat{\bf z})^2 =1,
\end{equation}
we have
\begin{eqnarray}
\hat{\bf n} \cdot (\hat{\bf n} \cdot {\bf M}_4 \cdot \hat{\bf z}) \cdot \hat{\bf z}
\xrightarrow{\rho\to 0}
4 \left( \frac{a^2k^2}{4a\rho} \right)^\frac{11}{2} \frac{z^2}{a^2}
\frac{1}{2} \frac{\pi}{2} 
\hspace{15mm} \nonumber \\ \times 
\left[ 1+ \left( 2\frac{z^2}{a^2} -1\right)
 (\hat{\bf n} \cdot \hat{\bf z})^2 \right], \hspace{8mm}
\end{eqnarray}
which matches the result in Eq.\,(\ref{m4-onax}).

Using Eqs.\,(\ref{m0co-840}), (\ref{m2co-840}),
and (\ref{m4co-840}), in Eq.\,(\ref{en-um024e}),
the Casimir-Polder interaction energy between
a completely anisotropically polarizable point atom
and a polarizable dielectric ring is given by
\begin{widetext}
\begin{eqnarray}
E&=& -\frac{\hbar c}{8\pi^2} \frac{\alpha_1\sigma_2}{a^6}
\Bigg[ 
(\hat{\bf n} \cdot \hat{\bf z})^2
\left\{ 13 \left( \frac{a^2k^2}{4a\rho} \right)^\frac{7}{2} \pi_7(k)
-56 \frac{z^2}{a^2} \left( \frac{a^2k^2}{4a\rho} \right)^\frac{9}{2} \pi_9(k)
+63 \frac{z^4}{a^4} \left( \frac{a^2k^2}{4a\rho} \right)^\frac{11}{2} \pi_{11}(k)
\right\}
\nonumber \\ &&
+63(\hat{\bf n} \cdot \hat{\bm\rho})^2
\left( \frac{a^2k^2}{4a\rho} \right)^\frac{11}{2} \frac{z^2}{a^2} \left\{
\left( 1 +\frac{\rho}{a} -\frac{2}{k^2} \right)^2 \pi_{11}(k)
+\frac{4}{k^2}\left( 1 +\frac{\rho}{a} -\frac{2}{k^2} \right) \pi_9(k)
+\frac{4}{k^4} \pi_7(k) \right\}
\nonumber \\ &&
+63(\hat{\bf n} \cdot \hat{\bm\phi})^2 
\left( \frac{a^2k^2}{4a\rho} \right)^\frac{11}{2} \frac{z^2}{a^2} \left\{
\frac{4}{k^2} \left( 1 -\frac{1}{k^2} \right) \pi_{11}(k)
-\frac{4}{k^2}\left( 1 -\frac{2}{k^2} \right) \pi_9(k)
-\frac{4}{k^4} \pi_7(k) \right\}
\nonumber \\ &&
-(\hat{\bf n} \cdot \hat{\bm\rho}) (\hat{\bf n} \cdot \hat{\bf z})
\bigg\{ 
56\left( \frac{a^2k^2}{4a\rho} \right)^\frac{9}{2} \frac{z}{a}
\left[\left( 1 +\frac{\rho}{a} -\frac{2}{k^2} \right) \pi_9(k)
+\frac{2}{k^2} \pi_7(k) \right] 
\nonumber \\ && \hspace{25mm}
-63\left( \frac{a^2k^2}{4a\rho} \right)^\frac{11}{2} 2\frac{z^3}{a^3}
\left[\left( 1 +\frac{\rho}{a} -\frac{2}{k^2} \right) \pi_{11}(k)
+\frac{2}{k^2} \pi_9(k) \right]
\bigg\} \Bigg].
\label{res-Ie-m3}
\end{eqnarray}
\end{widetext}
We emphasize that the electromagnetic interaction associated with the energy in
Eq.\,(\ref{res-Ie-m3}) for the configuration of an atom and a ring
is a manifestation of quantum fluctuations in vacuum.
These interactions are always characterized by $\hbar$ and in
the long-range Casimir-Polder limit are characterized by $\hbar c$,
as in Eq.\,(\ref{res-Ie-m3}).
This should be contrasted with short-range van der Waals
limit~\cite{Milton:2008lwv},
where the relativistic retardation effects are irrelevant
and have not been captured in the expression above.
The interaction energy in Eq.\,(\ref{res-Ie-m3}) is equal to $-E_0$
when the atom is exactly at the center of the ring and the
orientation of the polarizability of the atom is aligned
with respect to the axis of symmetry of the ring.
We will choose this value for the interaction energy, without the negative sign,
\begin{equation}
E_0=\frac{13}{16\pi} \frac{\hbar c}{a} \frac{\alpha_1\sigma_2}{a^5},
\end{equation}
to benchmark the scale of the interaction.
Observe that this value for the energy is also obtained
from Eq.\,(\ref{at-cen-48}) when $\hat{\bf n} \cdot \hat{\bf z}=1$.
To obtain a numerical estimate of this energy,
assume polarizabilities $\alpha_1$ and $\sigma_2$
to be on the order of the geometric size of a nanoparticle or an atom.
Further, assume that the radius of the ring is about ten times the size
of the nanoparticle. Then, we have $\alpha_1\sigma_2/a^5 \sim 10^{-5}$
and for $a=1$\,nm we evaluate $E_0$ to be on the order of meV.
Similarly, for $a=10$\,nm, the energy $E_0$ is 0.1\,meV.
For dielectrics, in contrast to conductors, the polarizabilities
(with units of volume) are smaller than the respective geometric sizes
and thus the associated energies are lower.
For example, if we assume the separation length of the polarized charges
in dielectrics to be forty percent smaller than in conductors,
an arbitrarily chosen value, it effectively lowers the interaction energy
by a factor of $(0.4)^5\sim 0.01$. Thus,
for $a=10$\,nm, the energy $E_0$ is about 1\,$\mu$eV.
In general, these interactions are significantly stronger
in conductors than dielectrics.

What about temperature effects? At room temperature, $kT=25$\,meV.
Thus, any sort of bonding achieved using the interaction energy
$E_0$ will be easily broken at room temperature.
One might also wonder whether these interactions
with their origins in quantum fluctuations can
be dwarfed by temperature fluctuations.
Quantum fluctuations dominate
over temperature fluctuations for 
\begin{equation}
kT \ll \frac{\hbar c}{a},
\end{equation}
where $a$ is a characteristic length in the system, say the
physical length associated with polarization of charges.
Thus, at room temperature, effects from quantum fluctuations
will be larger than those from temperature fluctuations
up to a humongous length scale of 10\,$\mu$m. This is consistent with
the experimental challenges faced in studying thermal contributions 
to Casimir effect~\cite{Lamoreaux:2011fct}.

\section{Energy landscape}
\label{sec-stab-ana}

The energy in Eq.\,(\ref{res-Ie-m3}) in units of $E_0$
is dependent on the position ${\bf r}$ of the atom
and the orientation of the polarizability $\hat{\bf n}$ of the atom,
see FIG.\,\ref{fig-aonr205}. More explicitly, the position
of atom is given in terms of cylindrical coordinates
($\rho$, $\phi$, $z$,)
and the respective cylindrical unit vectors,
$\hat{\bm\rho}$, $\hat{\bm\phi}$, and $\hat{\bf z}$,
refer Eq.\,(\ref{pacc5}),
and the polarizability is 
given in terms of the spherical polar coordinate $\theta_1$
and the spherical azimuth coordinate $\phi_1$, refer Eq.\,(\ref{pasc7}).
The dependence on the azimuth variables,
$\phi$ from position and $\phi_1$ from polarizability,
is of the form $\phi-\phi_1$.
Thus, we can write 
\begin{equation}
E= E({\bf r},\hat{\bf n}) = E(z,\rho,\phi-\phi_1,\theta_1).
\end{equation} 
The distances $z$ and $\rho$ are suitably expressed as
\begin{equation}
\frac{z}{a} \quad \text{and} \quad \frac{\rho}{a}
\end{equation}
in units of radius $a$ of the ring.
The energy in Eq.\,(\ref{res-Ie-m3})
has been organized in terms of four bilinear constructions
in the orientation dependence,
\begin{subequations}
\begin{eqnarray}
(\hat{\bf n} \cdot \hat{\bf z})^2 &=&\cos^2\theta_1,
\label{orpro59a} \\
(\hat{\bf n} \cdot \hat{\bm\rho})^2 &=&
\sin^2\theta_1\cos^2(\phi_1-\phi),
\label{orpro59b} \\
(\hat{\bf n} \cdot \hat{\bm\phi})^2 
&=&\sin^2\theta_1\sin^2(\phi_1-\phi), \\
(\hat{\bf n} \cdot \hat{\bm\rho})
(\hat{\bf n} \cdot \hat{\bf z})
&=&\cos\theta_1\sin\theta_1\cos(\phi_1-\phi),
\end{eqnarray}%
\label{orpro59}%
\end{subequations}%
constructed out of the projections of $\hat{\bf n}$.
Without loss of generality, we can choose
the $x$-axis to be oriented such that $\phi=0$. Thus,
the energy is parameterized as 
\begin{equation}
E = E(z,\rho,\phi_1,\theta_1).
\end{equation}

\begin{figure}
\includegraphics[width=8cm]{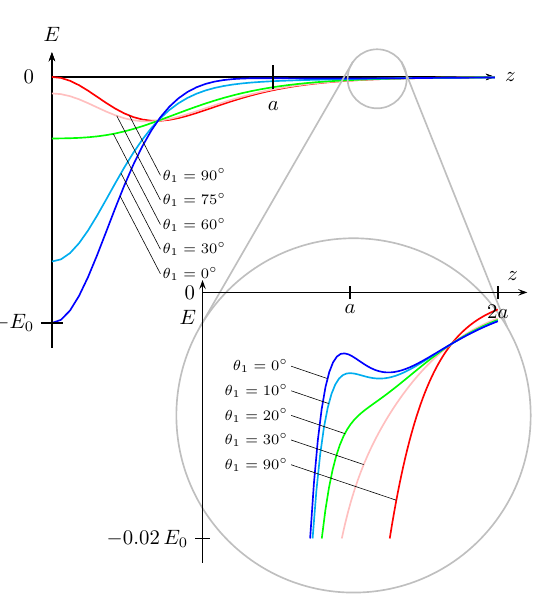}
\caption{Interaction energy in Eq.\,(\ref{res-Ie-m3}) for $\rho=0$
as a function of $z$.
For $\theta_1=0$, the energy has a minimum at $z=0$ and $z=\pm 1.259\,a$,
and maximum at $z=\pm 0.956\,a$.
For $\theta_1=90^\circ$, the energy has a maximum at $z=0$,
and minimum at $z=\pm 0.471\,a$.
}
\label{fig-en-ax-965}
\end{figure}%

Before we explore the energy landscape for features off the axis,
let us briefly recall the salient features of the energy
in Ref.~\cite{Marchetta:2020dap} when the unipolarizable atom is on the axis.
For $\rho\to 0$, the expression for energy in Eq.\,(\ref{res-Ie-m3})
leads to the expression we obtained in Eq.\,(\ref{enax-nz-e})
and reproduces the expression in Eq.\,(106)
of Ref.~\cite{Marchetta:2020dap}
and has been plotted here as FIG.~\ref{fig-en-ax-965}.
The energy is symmetric under $z\to-z$.
It was realized early on that in these configurations
the unipolarizable atom is unstable in the radial directions
everywhere on the symmetry axis.
In contrast, in the axial directions, the energy in FIG.~\ref{fig-en-ax-965}
exhibits maxima and minima, implying directional stability or instability 
in the axial directions.
If the ring were extended to a plane, there is no
maximum~\cite{Marchetta:2020dap}.
The points of minimum energy in FIG.~\ref{fig-en-ax-965}
correspond to stability
in the axial direction and instability in the radial direction,
and are effectively saddle points of equilibrium.
The points of maximum energy in FIG.~\ref{fig-en-ax-965}
correspond to instability in both axial and radial directions
and are thus unstable points.
We note that when the direction of polarizability is parallel
to the axis, for $\theta_1=0$, see FIG.~\ref{fig-en-ax-965}, we have
the center of ring,
\begin{subequations}
\begin{equation}
\rho=0, \quad z=0,
\end{equation}
as a saddle point of equilibrium;
\begin{equation}
\rho=0, \quad z=\pm \frac{a}{2}\sqrt{5+\frac{3}{\sqrt{5}}}
\approx \pm 1.3\,a,
\end{equation}
as another pair of shallow saddle points of equilibrium
on either sides of the ring; and
\begin{equation}
\rho=0, \quad z=\pm \frac{a}{2}\sqrt{5-\frac{3}{\sqrt{5}}}
\approx \pm 0.96\,a,
\end{equation}
\end{subequations}
as a pair of unstable points of equilibrium on either side of the ring.
When the direction of polarizability is perpendicular to the axis, 
for $\theta_1=90^\circ$, see FIG.~\ref{fig-en-ax-965}, we have
\begin{subequations}
\begin{equation}
\rho=0, \quad z=0,
\end{equation}
to be an unstable equilibrium point
and
\begin{equation}
\rho=0, \quad z=\pm a\frac{\sqrt{2}}{3}
\approx \pm 0.47\,a,
\end{equation}
\end{subequations}
as saddle points of equilibrium. The equilibrium points for arbitrary
orientations of the polarizability can be easily calculated when the atom
is confined on the axis, see Ref.~\cite{Marchetta:2020dap}.
The more general expression for energy in Eq.\,(\ref{res-Ie-m3})
allows us to explore the variation in energy in the 
vicinity of the above equilibrium points when we deviate in 
both radial and axial directions.
Here, we present plots for a wide range of orientations
of polarizability of the atom while restricting the analytical study near
these equilibrium points to the case of $\theta_1=0^\circ$ alone.
A comprehensive stability analysis and suitable applications
should be pursued elsewhere.

\begin{figure}
\includegraphics[width=8.5cm]{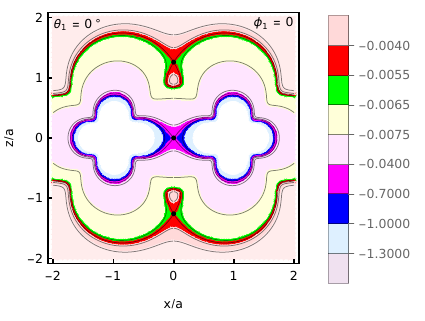}
\caption{Contours of equal energy in Eq.\,(\ref{th1=00ph100-as7}) 
plotted as a function of position $x/a$ and $z/a$ of polarizable atom.
The legend for the contours is energy measured in units of $E_0$,
that is, $E/E_0$.}
\label{fig-en-xyz-theta000-456}
\end{figure}%
\begin{figure}
\includegraphics[width=8.5cm]{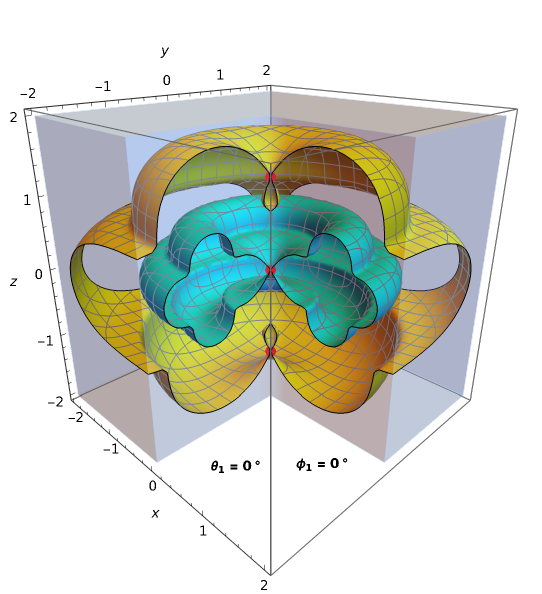 }
\caption{Equal energy surfaces for the energy in Eq.\,(\ref{th1=00ph100-as7})
plotted as a function of position, $x/a$, $y/a$, and $z/a$, of the polarizable atom.
The inner cyan surface corresponds to $E=-E_0$
and the outer yellow surface if for $E=-0.0065\,E_0$.
}
\label{fig-enSur-xyz-theta000-456}
\end{figure}%

When the polarization of the atom is aligned with the axis of symmetry
of the dielectric ring, that is, $\theta_1=0$,
out of the four projections in Eqs.\,(\ref{orpro59})
only the projection in Eq.\,(\ref{orpro59a}) is nonzero. 
This is an axially symmetric configuration with energy
\begin{eqnarray}
E(z,\rho,0,0)= -E_0\frac{2}{\pi} \Bigg[
\left( \frac{a^2k^2}{4a\rho} \right)^\frac{7}{2} \pi_7(k)
\hspace{25mm} \nonumber \\
-\frac{56}{13} \frac{z^2}{a^2} \left( \frac{a^2k^2}{4a\rho} \right)^\frac{9}{2} \pi_9(k) 
+\frac{63}{13} \frac{z^4}{a^4} \left( \frac{a^2k^2}{4a\rho} \right)^\frac{11}{2} \pi_{11}(k) 
\Bigg] \hspace{6mm}
\label{th1=00ph100-as7}
\end{eqnarray}
independent of $\theta_1$, $\phi_1$, and $\phi$.
Slices of surfaces of equal energy for the above energy in the
$x$-$z$ plane are illustrated as contours in FIG.~\ref{fig-en-xyz-theta000-456},
and this energy in the $z$ direction alone was
the blue curve ($\theta_1=0$) in FIG.~\ref{fig-en-ax-965}.
Quadrilobed-toroidal pattern in these surfaces is a signature of
the $\cos 2\theta$ dependence of Casimir-Polder interaction energy
in contrast to $\cos\theta$ dependence of dipole-dipole interaction energy. 
Two of the surfaces
of equal energy are shown in FIG.~\ref{fig-enSur-xyz-theta000-456}
in full glory in the complete three-dimensional space.
Lines normal to these surfaces of equal energy represent the
force on the polarizable atom, since it is a conservative force
and equal to the negative gradient of energy.
Further, the divergence of the force tells us about the stability at a point.
This is similar to electrostatics where negative gradient of electric potential
is electric field and divergence of electric field
at a point is a measure of electric charge density. 
In electrostatics, charge density is a source or sink of electric field.
Similarly, Laplacian of energy (in contrast to electric potential)
is a source or sink of force, and thus a measure of stability.
Thus, in FIG.~\ref{fig-en-xyz-theta000-456} and FIG.~\ref{fig-enSur-xyz-theta000-456},
points where the energy surfaces are locally concentric correspond to
stable or unstable points and points where the energy surfaces gets
pinched and appear as intersecting contours in their cross sections,
such as in FIG.~\ref{fig-en-xyz-theta000-456}, are saddle points of equilibrium.
Thus, we decipher that the energy in Eq.\,(\ref{th1=00ph100-as7})
has three equilibrium points on the $z$ axis, one at the origin
and one each on either side of the ring on $z$ axis at $z_0=\pm 1.3\,a$.
We also learn that the unstable points of equilibrium at $z_0=\pm 0.96\,a$
appear as hanging blobs in FIG.~\ref{fig-enSur-xyz-theta000-456}.
The force on the atom placed at each of these equilibrium points is zero.
We deduce that $z_0=\pm 0.96\,a$ is an unstable point of equilibrium
because the concentric surfaces are lower in energy going away from this point. 
For the three saddle points out of the five equilibrium points, energy decreases
in radial directions and energy increases in axial directions.
This will be manifested as an attractive force
(toward the equilibrium point) in the direction of $z$
and a repulsive force (away from the equilibrium point) in radial directions.

To gain analytical insight into the structure of the energy surface
in the neighborhood of the equilibrium points
we use the expansion, using Eq.\,(\ref{k2def-48}),
\begin{equation}
\frac{a^2k^2}{4a\rho} =1-2\frac{\rho}{a}
+3\frac{\rho^2}{a^2} -\frac{z^2}{a^2} +\dots,
\end{equation}
in the series expansions for $\pi_n(k)$ in Eqs.\,(\ref{piN_powerSeriesExpansions})
to obtain
\begin{equation}
\left( \frac{a^2k^2}{4a\rho} \right)^\frac{7}{2} \pi_7(k)
=\frac{\pi}{2} \left[1+\frac{49}{4} \frac{\rho^2}{a^2}
-\frac{7}{2} \frac{z^2}{a^2} +\ldots \right],
\end{equation}
and
\begin{subequations}
\begin{eqnarray}
\frac{z^2}{a^2} \left( \frac{a^2k^2}{4a\rho} \right)^\frac{9}{2} \pi_9(k)
&=& \frac{z^2}{a^2} \frac{\pi}{2} +\ldots, \\
\frac{z^4}{a^4} \left( \frac{a^2k^2}{4a\rho} \right)^\frac{11}{2} \pi_{11}(k)
&=& 0 +\ldots,
\end{eqnarray}
\end{subequations}
which yields the form of energy in the vicinity of the equilibrium 
point at the origin to be
\begin{equation}
z_0=0: \hspace{3mm} E = -E_0 \left[ 1 +\frac{49}{4}\frac{\rho^2}{a^2}
-\frac{203}{26}\frac{z^2}{a^2} +\ldots \right].
\end{equation}
The overall negative coefficient for $\rho^2$ implies a negative
second derivative; thus, instability in
radial directions and a positive coefficient for $z^2$
implies stability in axial directions.

\begin{figure*}%
\includegraphics[width=5.9cm]{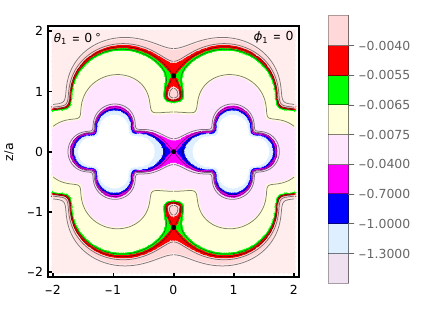}
\includegraphics[width=5.4cm]{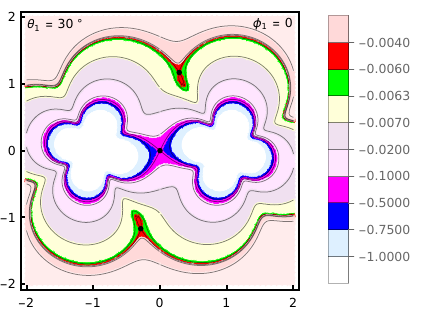}
\includegraphics[width=5.4cm]{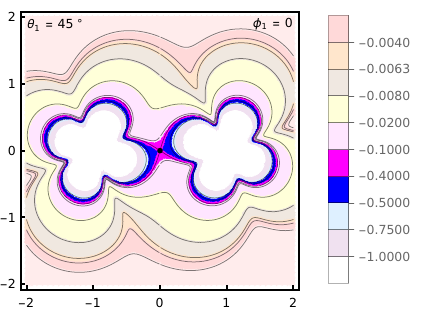}
\includegraphics[width=5.9cm]{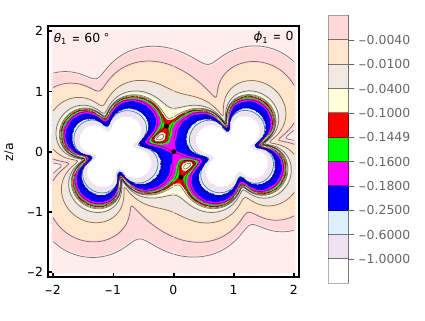}
\includegraphics[width=5.4cm]{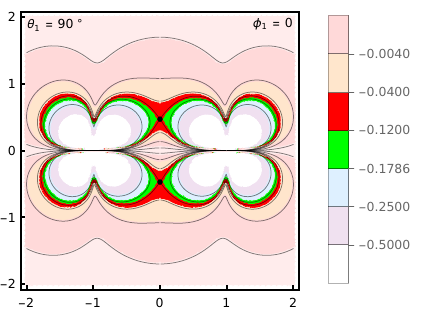}
\includegraphics[width=5.4cm]{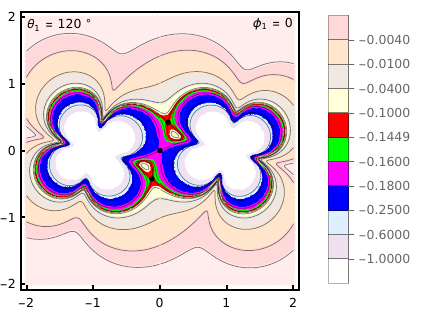}
\includegraphics[width=5.9cm]{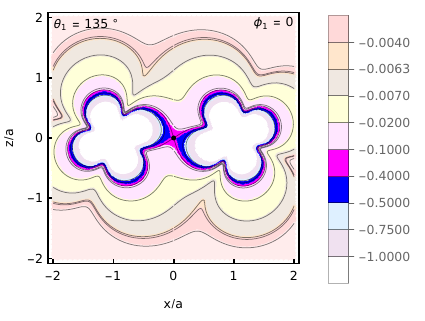}
\includegraphics[width=5.4cm]{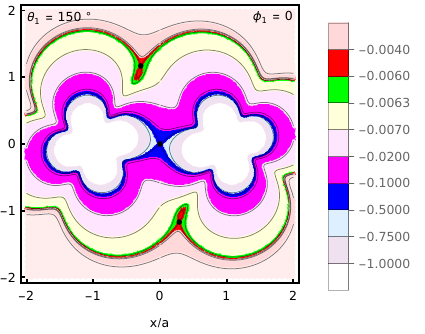}
\includegraphics[width=5.4cm]{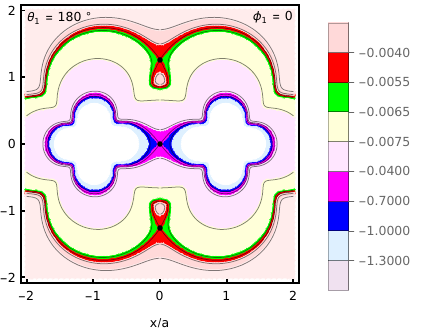}
\caption{
Each of the frames show cross sections of surfaces of equal energy
in Eq.\,(\ref{res-Ie-m3}) as contours in the region of space surrounding
the unipolarizable dielectric ring of radius $a$ as a function of
position $\rho$ and $z$ of the unipolarizable atom.
The cross section in each frame is the $x$-$z$ plane.
Each of the nine frames correspond to a fixed azimuth angle $\phi_1=0^\circ$
of the polarizability of atom, and the variation in the frames
is in the orientation angle $\theta_1$ of the polarizability.
The frame on the top left, for $\theta_1=0^\circ$, has three saddle points
of equilibrium described by black dots, one at the origin and two (one on
each side of ring) on the $z$ axis at $z_0=\pm 1.3\,a$.
These three equilibrium points drift away from the $z$ axis
as we change the orientation of the polarizability of the atom
in the following frames
for angles $\theta_1= 0^\circ, 30^\circ, 45^\circ, 60^\circ, 90^\circ,$
and then for $\theta_1= 180^\circ, 150^\circ, 135^\circ, 120^\circ, 90^\circ$.
The legend for the contours is energy measured in units
of $E_0$, that is, $E/E_0$.
The contours show an inversion symmetry about the origin
between orientations $90^\circ<\theta_1<180^\circ$
and $0^\circ<\theta_1<90^\circ$.
}%
\label{fig-en-xz-933}%
\end{figure*}%
\begin{figure*}%
\includegraphics[width=5.9cm]{figures/epsfiles/energy-contour3D-versus-x-y-z-for-theta000-phi000.pdf}
\includegraphics[width=5.9cm]{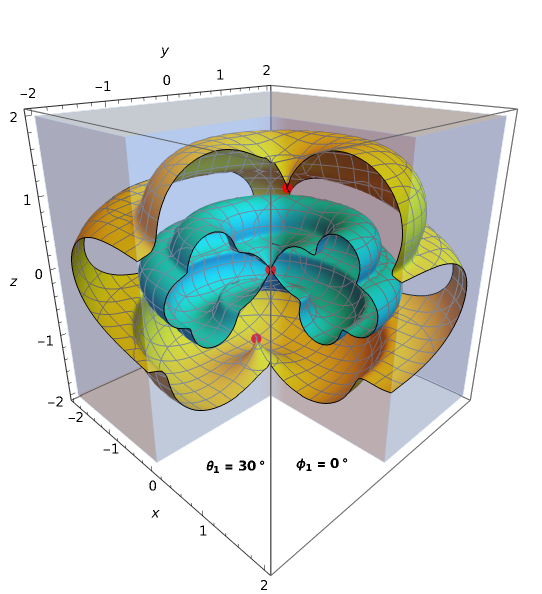}
\includegraphics[width=5.9cm]{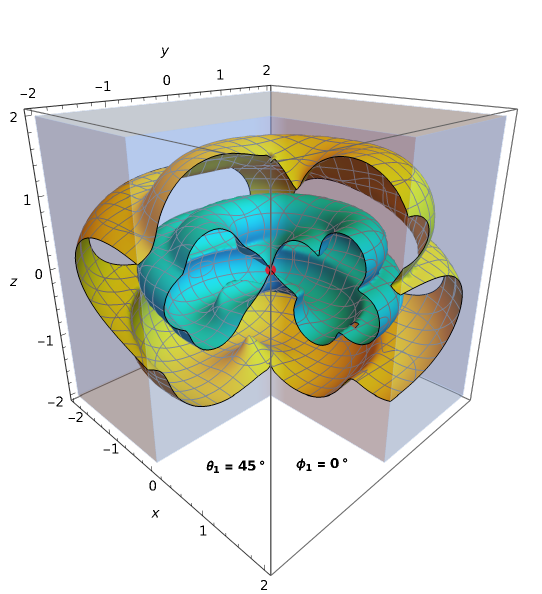}
\includegraphics[width=5.9cm]{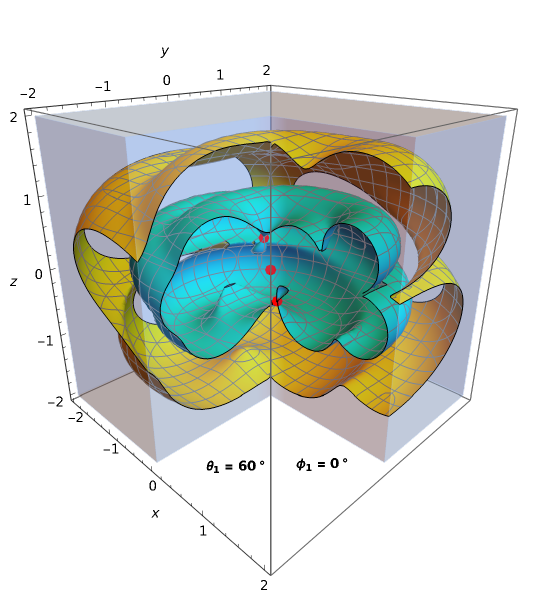}
\includegraphics[width=5.9cm]{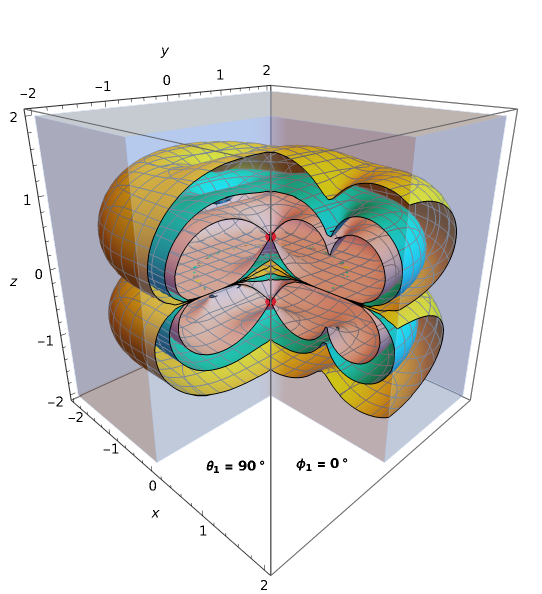}
\includegraphics[width=5.9cm]{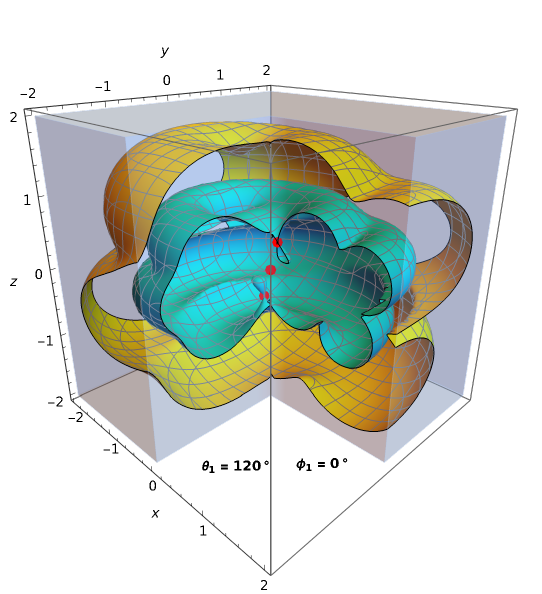}
\includegraphics[width=5.9cm]{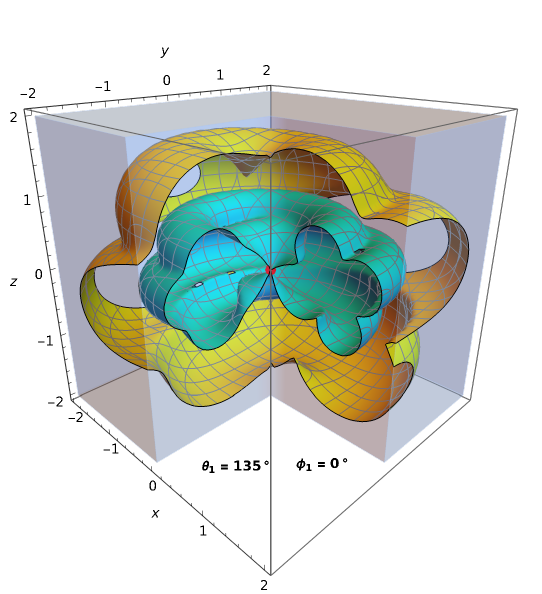}
\includegraphics[width=5.9cm]{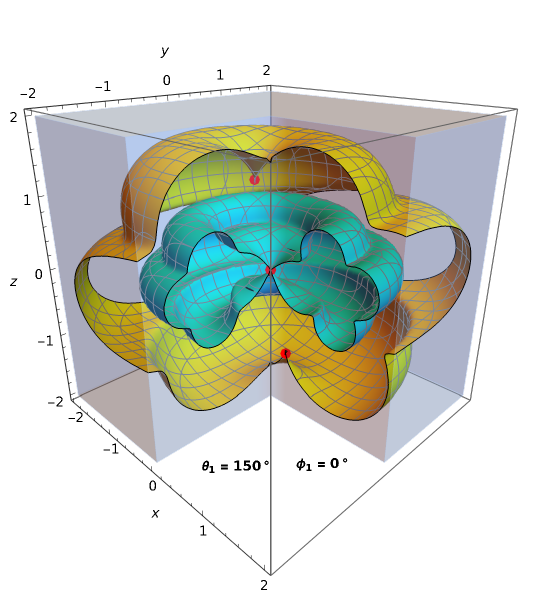}
\includegraphics[width=5.9cm]{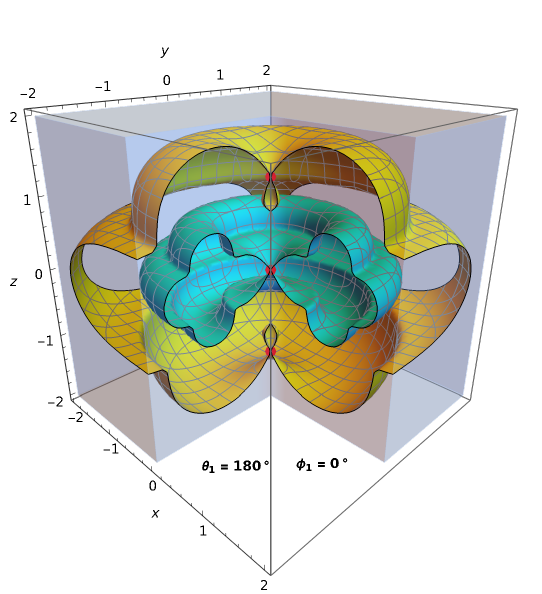}
\caption{
Each of the frames show two surfaces of equal energy
in Eq.\,(\ref{res-Ie-m3}) in the region of space surrounding
the unipolarizable dielectric ring of radius $a$
as a function of position of the unipolarizable atom.
The variables on the axes, $x$, $y$, and $z$, represent the
position of the polarizable atom and are in multiples of radius $a=1$ of ring.
Each of the nine frames correspond to a fixed azimuth angle $\phi_1=0^\circ$
of the polarizability of atom, and the variation in the frames
is in the orientation angle $\theta_1$ of the polarizability.
The equilibrium points are described by red dots. The center of the ring
is a saddle point of equilibrium except when $\theta_1=90^\circ$.
Orientations for $90^\circ<\theta_1<180^\circ$ is identical to
$0^\circ<\theta_1<90^\circ$ with an inversion symmetry about the origin.
The (outer) yellow colored surface for each $\theta_1$ is the energy surface that
passes through two (one on each side of ring) saddle points of equilibrium,
with energies $E=-E_0(0.0065,0.0063,0.0063,0.004,0.004)$
for angles $\theta_1= 0^\circ, 30^\circ, 45^\circ, 60^\circ, 90^\circ,$
and the same for $\theta_1= 180^\circ, 150^\circ, 135^\circ, 120^\circ, 90^\circ,$.
The (inner) cyan colored surfaces are the energy surfaces that contain
the saddle points of equilibrium closer to the ring
with energies $E=-E_0(1,0.75,0.5,0.16,0.178)$
for angles $\theta_1= 0^\circ, 30^\circ, 45^\circ, 60^\circ, 90^\circ,$
and the same for $\theta_1= 180^\circ, 150^\circ, 135^\circ, 120^\circ, 90^\circ$.
}%
\label{fig-en-xyz-933}%
\end{figure*}%

Similar series expansions in the neighborhood of
the five equilibrium points discussed above,
keeping until quadratic terms, are
\begin{subequations}
\begin{align}
\frac{z_0}{a}&=0: & \frac{E}{E_0}
&= -1 -12\frac{\rho^2}{a^2} +7.8 \frac{\tilde z^2}{a^2}, \\
\frac{z_0}{a}&=\pm 0.96: & \frac{E}{E_0}
&= -0.0050 -0.11\frac{\rho^2}{a^2} -0.097 \frac{\tilde z^2}{a^2}, \\
\frac{z_0}{a}&=\pm 1.3: & \frac{E}{E_0}
&= -0.0065 -0.61\frac{\rho^2}{a^2} +0.024 \frac{\tilde z^2}{a^2},
\end{align}
\end{subequations}
where
\begin{equation}
\tilde z = z-z_0,
\end{equation}
are deviations from respective equilibrium points $z_0$ given in units of ring radius $a$. 
The relative signs and values of the coefficients of $\rho^2$ and
$z^2$ terms reveal the nature of the equilibrium points. Thus,
we infer, again, that $z_0=\pm 1.3\,a$ and $z_0=0$ are saddle points
of equilibrium and $z_0=\pm 0.96\,a$ are unstable points of equilibrium.
The force on the polarizable atom in the vicinity of these
equilibrium points is evaluated using
\begin{equation}
{\bf F} = -{\bm\nabla} E,
\end{equation}
in terms of $F_0=E_0/a$ as
\begin{subequations}
\begin{align}
\frac{z_0}{a}&=0: & {\bf F} &= F_0\left[ 
24\frac{\rho}{a} \hat{\bm\rho} -16 \frac{\tilde z}{a} \hat{\bf z} \right], \\
\frac{z_0}{a}&=\pm 0.96: & {\bf F} &= F_0 \left[
0.22\frac{\rho}{a} \hat{\bm\rho} +0.19 \frac{\tilde z}{a} \hat{\bf z} \right], \\
\frac{z_0}{a}&=\pm 1.3: & {\bf F} &= F_0 \left[ 
0.12\frac{\rho}{a} \hat{\bm\rho} -0.048 \frac{\tilde z}{a} \hat{\bf z} \right],
\end{align}
\end{subequations}
each of which is easily verified to be zero at the respective equilibrium points.
It is known that the stability or instability at these equilibrium points is
not conclusively measured by the divergence of force, the Laplacian
of the negative of energy~\cite{Teller:1937spc}.
The divergence of force at the equilibrium points are
\begin{subequations}
\begin{align}
\frac{z_0}{a}&=0: & \frac{{\bm\nabla} \cdot {\bf F}}{(F_0/a)}
&= (48 -16)= 32, \\
\frac{z_0}{a}&=\pm 0.96: & \frac{{\bm\nabla} \cdot {\bf F}}{(F_0/a)} 
&= (0.44 +0.19)= 0.63, \\
\frac{z_0}{a}&=\pm 1.3: & \frac{{\bm\nabla} \cdot {\bf F}}{(F_0/a)} 
&= (0.24 -0.048)= 0.19, 
\end{align}
\end{subequations}
where a multiplicative factor of 2 in the $\rho$ directions is due
to the contribution from the angular dependence in $\hat{\bm\rho}$. 
That is, ${\bm\nabla}\cdot (\rho\hat{\bm\rho})=2$.
Thus, deducing that $z_0=\pm 0.96\,a$ is an unstable equilibrium point
in this manner is not conclusive but consistent with the inference
made earlier using graphical analysis.
For completeness, we also verify that
\begin{equation}
{\bm\nabla} \times {\bf F} = 0 
\end{equation}%
at each of the above equilibrium points.
We also point out that the yellow outer surface
in FIG.~\ref{fig-enSur-xyz-theta000-456}
on which four of the five equilibrium points reside are
about hundred times smaller in energy
relative to the cyan inner surface.
Thus, the valley constituting the saddle points are shallow
and probably not very easily accessible for applications.
Nevertheless, the features in the energy landscape are insightful.

\begin{figure}
\includegraphics{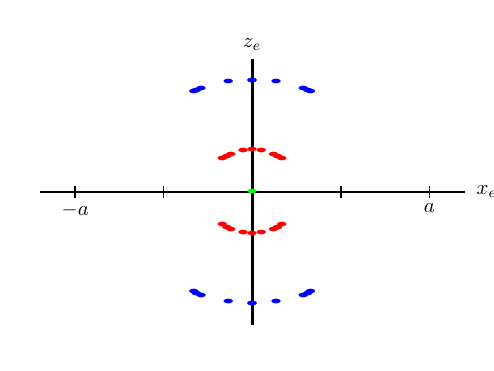} 
\caption{Drift of position $(x_e, y_e, z_e)$ of equilibrium points on 
equal energy surfaces for the energy in Eq.\,(\ref{th1=00ph100-as7})
as a function of the orientation of the polarization
$(\theta_1,\phi)$ of the unipolarizable atom.
The blue dots are graphical representations of the data in
Table~\ref{table-drift-ofEqP-1p3-501}
that are absent for $33^\circ <\theta_1 < 147^\circ$,
and red dots are graphical representations of the data in
Table~\ref{table-drift-ofEqP-0p47-501}
that are absent for $\theta_1<55^\circ$ and $125^\circ<\theta_1$.
The equilibrium point at origin does not drift.
}
\label{fig-drift-ofEqP-501}
\end{figure}

We can extend the analysis of each of the equilibrium points
for orientation $\theta_1=0^\circ$ as we change the orientation.
This is summarized adequately by showing the drift of these
equilibrium points as a function of orientation angle.
This has been summarized in
Tables~\ref{table-drift-ofEqP-1p3-501} and \ref{table-drift-ofEqP-0p47-501}.
The equilibrium points are of course just the highlights of
the energy surfaces in FIGs.~\ref{fig-en-xz-933} and \ref{fig-en-xyz-933}.
The contours in FIGs.~\ref{fig-en-xz-933}
are curves that represent cross sections in the $x$-$z$ plane of
equal energy surfaces.
The complete surface (which the curves in FIGs.~\ref{fig-en-xz-933}
are part of) are shown in FIG.~\ref{fig-en-xyz-933}.
The specific energy surfaces we have displayed for 
each of the orientations are those that contain equilibrium points,
or are in close vicinity of these surfaces.
This graphical display is crucial for investigating the stability
of the equilibrium points. We mention that concrete analytical methods
to investigate the stability of equilibrium points in three dimensions
are lacking and graphical analysis is the tool of choice.
The basics of graphical stability analysis is to look for concentric
surfaces around an equilibrium point
representing stability or instability in all three directions
and for conical intersections
in the energy surfaces representing instability in at least one of the
three directions.
This is similar to the analysis described in Ref.\,\cite{Teller:1937spc},
though in a different context.
The intersections displayed in FIGs.~\ref{fig-en-xz-933} and \ref{fig-en-xyz-933}
were determined by manually searching for the specific energy.
This was a tedious affair, but can be easily automated if it
has to be done at a bigger scale.
Data for each frame in FIG.~\ref{fig-en-xz-933} took about five minutes
to be generated on a typical PC, and data for each frame in
FIG.~\ref{fig-en-xyz-933} was generated in about five hours.
The drift of the equilibrium points as a function of orientation angle,
in Tables~\ref{table-drift-ofEqP-1p3-501} and \ref{table-drift-ofEqP-0p47-501},
has been illustrated in FIG.~\ref{fig-drift-ofEqP-501}.

\begin{table}
\begin{tabular}{cc cc c}
\hline
$(\theta_1,\phi_1)$ &~~~& $(x_e, y_e, z_e)$ &~~~& $E$ \\
\hline
$(0^\circ,0^\circ)$   && $(0.00\,a, 0, 1.3\,a)$ && $-0.0065\,E_0$ \\
$(15^\circ, 0^\circ)$ && $(0.14\,a, 0, 1.2\,a)$ && $-0.0064\,E_0$ \\
$(30^\circ, 0^\circ)$ && $(0.29\,a, 0, 1.2\,a)$ && $-0.0063\,E_0$ \\
$(32^\circ, 0^\circ)$ && $(0.32\,a, 0, 1.2\,a)$ && $-0.0063\,E_0$ \\
$(33^\circ, 0^\circ)$ && $(0.33\,a, 0, 1.1\,a)$ && $-0.0063\,E_0$ \\
\hline
$(147^\circ, 0^\circ)$ && $(-0.33\,a, 0, 1.1\,a)$ && $-0.0063\,E_0$ \\
$(148^\circ, 0^\circ)$ && $(-0.32\,a, 0, 1.2\,a)$ && $-0.0063\,E_0$ \\
$(150^\circ, 0^\circ)$ && $(-0.29\,a, 0, 1.2\,a)$ && $-0.0063\,E_0$ \\
$(165^\circ, 0^\circ)$ && $(-0.14\,a, 0, 1.2\,a)$ && $-0.0064\,E_0$ \\
$(180^\circ,0^\circ)$   && $(0.00\,a, 0, 1.3\,a)$ && $-0.0065\,E_0$ \\
\hline
\end{tabular}
\caption{Position of equilibrium point at $(x_e, y_e, z_e)$
that originates at $(0, 0, \pm1.3\,a)$ on
equal energy surfaces for the energy $E$ in Eq.\,(\ref{th1=00ph100-as7})
as a function of the orientation of the polarization
$(\theta_1,\phi_1)$ of the unipolarizable atom.
This equilibrium point is absent for $33^\circ <\theta_1 < 147^\circ$.
}
\label{table-drift-ofEqP-1p3-501}
\end{table}%

\begin{table}
\begin{tabular}{cc cc c}
\hline
$(\theta_1,\phi_1)$ &~~~& $(x_e, y_e, z_e)$ &~~~& $E$ \\
\hline
$(55^\circ, 0^\circ)$ && $(-0.17\,a, 0, 0.37\,a)$ && $-0.16\,E_0$ \\
$(57^\circ, 0^\circ)$ && $(-0.14\,a, 0, 0.40\,a)$ && $-0.16\,E_0$ \\
$(60^\circ, 0^\circ)$ && $(-0.12\,a, 0, 0.42\,a)$ && $-0.16\,E_0$ \\
$(75^\circ, 0^\circ)$ && $(-0.05\,a, 0, 0.46\,a)$ && $-0.17\,E_0$ \\
$(90^\circ, 0^\circ)$ && $(0, 0, 0.47\,a)$        && $-0.18\,E_0$ \\
$(105^\circ, 0^\circ)$ && $(0.05\,a, 0, 0.46\,a)$ && $-0.17\,E_0$ \\
$(120^\circ, 0^\circ)$ && $(0.12\,a, 0, 0.42\,a)$ && $-0.16\,E_0$ \\
$(123^\circ, 0^\circ)$ && $(0.14\,a, 0, 0.40\,a)$ && $-0.16\,E_0$ \\
$(125^\circ, 0^\circ)$ && $(0.17\,a, 0, 0.37\,a)$ && $-0.16\,E_0$ \\
\hline
\end{tabular}
\caption{Position of equilibrium point at $(x_e, y_e, z_e)$ 
that originates at $(0, 0, \pm0.47\,a)$ on
equal energy surfaces for the energy $E$ in Eq.\,(\ref{th1=00ph100-as7})
as a function of the orientation of the polarization 
$(\theta_1,\phi_1)$ of the unipolarizable atom.
This equilibrium point is absent for $\theta_1<55^\circ$ and $125^\circ<\theta_1$.
}
\label{table-drift-ofEqP-0p47-501}
\end{table}%

When the polarization of the atom is perpendicular to the axis of symmetry
of the dielectric ring, that is, $\theta_1=90^\circ$,
and if $\phi_1-\phi=0$, then
out of the four projections in Eqs.\,(\ref{orpro59})
only the projection in Eq.\,(\ref{orpro59b}) is nonzero.
The energy for this configuration is
\begin{eqnarray}
E(z,\rho,90^\circ,0)= -E_0\frac{63}{13}
\frac{z^2}{a^2} \left( \frac{a^2k^2}{4a\rho} \right)^\frac{11}{2}
\frac{2}{\pi} \Bigg[\frac{4}{k^4} \pi_7(k)
\hspace{10mm} \nonumber \\
+\frac{4}{k^2}\left( 1 +\frac{\rho}{a} -\frac{2}{k^2} \right) \pi_9(k)
+\left( 1 +\frac{\rho}{a} -\frac{2}{k^2} \right)^2 \pi_{11}(k)
\Bigg]. \hspace{7mm}
\label{th1=90ph100-as7}
\end{eqnarray}
For finding series expansions around equilibrium points, it is 
convenient to rewrite it in the form
\begin{eqnarray}
E(z,\rho,90^\circ,0) &=& -E_0\frac{63}{13}
\frac{z^2}{a^2} \left( \frac{a^2k^2}{4a\rho} \right)^\frac{11}{2}
\frac{2}{\pi}
\hspace{10mm} \nonumber \\
&& \times \Bigg[
c_1(k) +2 \frac{\rho}{a} c_2(k)
+\frac{\rho^2}{a^2} \pi_{11}(k)
\Bigg], \hspace{7mm}
\label{th1=90ph100-ass46}
\end{eqnarray}
where $c_1(k)$ is given by the expression in
Eq.\,(\ref{nhDrh-se91})
and $c_2(k)$ is given by the expression in
Eq.\,(\ref{nhDrh-te912}).
The expansions of energy in the vicinity of the three
equilibrium points are
\begin{subequations}
\begin{align}
\frac{z_0}{a}&=0: & \frac{E}{E_0}
&= 0 -2.4 \frac{\tilde z^2}{a^2}
-68 \frac{\rho^2}{a^2}\frac{\tilde z^2}{a^2}, \\
\frac{z_0}{a}&=\pm 0.47: & \frac{E}{E_0}
&= -0.18 +1.3 \frac{\tilde z^2}{a^2}
-2.8 \frac{\rho^2}{a^2}\frac{\tilde z^2}{a^2}.
\end{align}
\end{subequations}

An animation illustrating the variation in the energy surfaces
including the drift of equilibrium points as a function
of polarization of the unipolarizable atom is available in  
Ref.\,\cite{Shajesh2025AnimAR} as a supplementary material.
The geometrical shape of energy surfaces can be understood
in the following manner.
The equal energy surfaces of a point electric charge are concentric spheres.
The equal energy surfaces of point electric dipole are bilobal,
each lobe with opposite sign of energy.
The equal energy surfaces of a polarizable particle are quadrilobal,
that is, has four lobes, with all four lobes having the same sign of energy.
A polarizable ring has equal energy surface in the shape of a toroid
with the cross section of the channel constituting the toroid
having the shape of a quadrilobe.
If we imagine the quadrilobe-toroidal-shaped energy surface to be
constructed by stretching and compressing the energy surface of a
line of polarizable particles, we can imagine the surfaces to be squeezed
near the center of the ring.
This energy surface has to coexist by topologically transforming itself
to accommodate the quadrilobe of a polarizable atom.
This interpretation is consistent with
the energy surfaces in FIGs.~\ref{fig-en-xz-933} and \ref{fig-en-xyz-933}.

\section{Conclusion and Outlook}
\label{sec-conclu}

We presented an exact expression for the
Casimir-Polder interaction energy between a unipolarizable atom
and a unipolarizable dielectric ring in terms of elliptic integrals
in Eq.\,(\ref{res-Ie-m3}).
The qualitative features of the results associated to the
equilibrium points on the $z$ axis were already known in the literature.
The quantitative analysis of stability at the equilibrium points,
both on the axis and off the axis, are new. The exact analytic
expression for energy allows us to find quadratic approximation
of the same in the vicinity of equilibrium points,
which in turn allows us to investigate the nature of stability at
these equilibrium points. Contours of equal energy surfaces,
not to be confused with electric equipotential surfaces,
are used to analyze stability.
Conical intersections in energy surfaces represent saddle points
and concentric surfaces imply stable or unstable points.
This procedure will be useful to investigate
feasibility of designs for stabilizing and thus trapping an atom
in interactions based on vacuum fluctuations.
We have confined the stability analysis to directions of polarization listed in
Tables~\ref{table-drift-ofEqP-1p3-501} and \ref{table-drift-ofEqP-0p47-501}.
The set of equilibrium points reported are exhaustive,
but are limited by the manual graphical analysis used for the process.
Thus, to be cautious, we hold back, and claim the analysis here
to be only a step towards a comprehensive stability analysis.
Another limitation of our analysis is uniaxial polarizability,
a constraint we imposed for simplicity.
A complete study should further include what happens
to the instability when the restriction to uniaxial polarizability
is relaxed.

Earnshaw theorem in electrostatics states that stable equilibrium
points are not possible in static configurations.
Thus, surfaces of equal energy for a static charge configuration
will not allow for concentric surfaces in a region such that the inner
surfaces are lower in energy relative to the outer.
For example, for the case of a point magnet interacting with
a ring magnet the equilibrium points are always saddle points;
that is, it is unstable in at least one direction out of three
and the intersections in energy surfaces are
conical~\cite{Warnakulasooriya2023mrm}. Earnshaw theorem
is well known to be not applicable for non-static configurations.
For example, a spinning magnet can levitate above a ring magnet,
like a Levitron$^\text{TM}$~\cite{Berry:1996sta}. Similarly,
electromagnets with alternating current can levitate magnets.
In light of these examples we should not expect Earnshaw
theorem to hold for interactions manifested by vacuum fluctuations
because they are inherently dynamic.
Thus, stability analysis in Casimir-Polder interactions
with a likelihood for a violation of Earnshaw theorem is warranted.
Earnshaw theorem has been discussed for polarizable atoms
in Ref.~\cite{Ashkin1978prr}
and for vacuum-fluctuations-induced interactions in
Ref.~\cite{Rahi2010fif}.
Unlike the case of a point magnet interacting with a ring
magnet~\cite{Warnakulasooriya2023mrm}
where we only see conical intersections in energy surfaces,
our analysis here suggests that we can find equilibrium
points with locally concentric energy surfaces around it
in Casimir-Polder interactions. See, for example,
$\theta_1=0$ case around $z=\pm 0.96\,a$ and $\rho=0$,
which appears as a hanging blob in FIG.~\ref{fig-en-xyz-theta000-456}
with concentric surfaces inside the blob.
These equilibrium points lead to instability in all three directions.
This observation leads to the question of whether it is
possible to have such concentric surfaces such that the 
equilibrium points lead to stability. 
This motivates us to inquire whether there are subtle differences
in the statement of Earnshaw theorem in the context of
vacuum fluctuations. 

We end our discussion by presenting a speculative outlook,
proposing to influence the nature of an equilibrium point
by distorting the polarizability.
In molecular chemistry, it is known that equal energy surfaces for
two atoms do not intersect if the interatomic distance is the
only parameter in the energy~\cite{Wigner:1929pae}.
However, if the energy depends on additional parameters,
like the orientation of the dipole moments of the atoms,
then the energy surfaces accommodate
conical intersections~\cite{Teller:1937spc}.
It has been further shown that the spatial degeneracy in energy
causes the molecules to distort,
the Jahn-Teller effect~\cite{Teller:1937do1}.
Extending these ideas from molecular chemistry to 
Casimir-Polder interactions is promising. 
We could inquire whether a distortion in the polarizability
of the ring could inherently change the nature of equilibrium points.
A distortion that is presumably tractable
is a configuration where the polarizability of the ring,
$\sigma_2 \, \hat{\bf m}\hat{\bf m}$,
is nonuniform along the ring,
\begin{equation}
\hat{\bf m} = \cos\phi \,\hat{\bf z} +\sin\phi \,\hat{\bm\rho}.
\end{equation}  
Qualitatively, we can imagine a quadrilobed-toroidal pattern,
which is a useful picturization of a polarizable atom,
to be twisting and turning once as it loops and forms a ring.
The effect this azimuthal asymmetry has on the interaction
energy and the associated equilibrium points
would suggest a Jahn-Teller effect in Casimir-Polder interactions
at macroscopic scale in contrast to molecular scale.

\section{acknowledgments}
We thank Dinuka Gallaba and Duston Wetzel for collaborative assistance.
KVS thanks Bumsu Lee and Punit Kohli at SIUC for discussions.

\bibliography{biblio/b20131003-casimir-top,%
biblio/b20210330-levi,biblio/b20230817-niranjan}

\end{document}